%% file: urban_rural_mob_ineff.tex
\documentclass[fleqn,10pt]{wlscirep}
\usepackage[utf8]{inputenc}
\usepackage{esvect}
\usepackage[T1]{fontenc}
\usepackage{xr-hyper}
\usepackage{hyperref}

\usepackage{fancyhdr} 
\title{Understanding Urban-Rural Disparities in Mobility Inefficiency for Colombia, Mexico, and India}

\author[1,+]{Nandini Iyer}
\author[2]{Massimiliano Luca}
\author[1,3,*]{Riccardo {Di Clemente}}
\affil[1]{Complex Connections Lab, Network Science Institute, Northeastern University London, UK.}
\affil[2]{Bruno Kessler Foundation, Via Sommarive 18, 38123 Povo (TN), Italy.}
\affil[3]{Institute for Scientific Interchange, 10126 Turin, Italy.}

\affil[+]{nandini.iyer@nulondon.ac.uk}
\affil[*]{riccardo.diclemente@nulondon.ac.uk}

\newcommand\norm[1]{\left\lVert#1\right\rVert}

\begin{abstract}

Rural and urban areas exhibit distinct mobility patterns, yet a systematic understanding of how these trends differ across regions and contexts remains underexplored. By using origin-destination matrices from Location-Based Services data in Colombia, India, and Mexico, we delineate urban and rural boundaries through network percolation, reducing reliance on conventional urbanisation metrics tied to the built environment. We gauge mobility dynamics across regions developing a measure for routing inefficiency, which measures how much longer empirical trips are than their optimal shortest path. Our findings reveal that rural areas experience greater inefficiencies, particularly for longer trips made later in the day. At the urban level, we determine the misalignment between urban mobility efficiency and public transit accessibility, by measuring the difference between their respective vector fields. We observe that most cities experience misalignment during regular commuting hours, with Colombian cities exhibiting particularly high alignment. Meanwhile, mobility inefficiency in rural areas are associated with their orientation around their most proximate city. City-level analyses uncover disparities in the functions of rural and urban areas, with significant variations between weekdays and weekends, reflecting distinct roles in commuting and access to services. These findings highlight the importance of tailored, context-sensitive approaches to improving connectivity and reducing disparities. This study offers new insights into the spatial and temporal dynamics of mobility inefficiency, contributing to equitable regional planning and sustainable mobility solutions.
\end{abstract}
\begin{document}

\flushbottom
\maketitle

\thispagestyle{empty}

\section*{Introduction}

As rates of urbanisation increase, it has become increasingly important to adopt data-driven approaches to promptly address challenges that arise in both urban and rural areas, such as monitoring economic development \cite{pappalardo2015using,wang2024infrequent, chetty2014land, chetty2016effects, chetty2022social}, accessibility \cite{vitale2020improving,graells2021city,abbiasov202415} and, inequalities \cite{barbosa2021uncovering, de2023quantifying, testi2024big, moro2021mobility, fan2023diversity}.
Yet, the process of urbanisation varies across countries and cultures \cite{edensor2012introduction,robinson2013ordinary}. Countries in the Global South tend to exhibit urbanisation through the growth of small urban settlements \cite{randolph2024urbanization}, while countries in the Global North do so through population density increases in megacities \cite{castells2010globalisation}. Despite the variability in urbanisation, investments in mobility infrastructure tend to be concentrated in larger urban areas \cite{moseley2023accessibility}.
Meanwhile, smaller urban areas and rural towns face mobility challenges due to limited transportation options and poorer access to urban hubs, often leading to a higher dependence on private vehicles \cite{shergold2012rural,gray2001car}. These disparities in accessibility and infrastructure can hinder economic growth \cite{cattaneo2021global}, exacerbate inequalities in healthcare \cite{ma2021urban} and result in disproportionate consequences of climate change \cite{fang2020cascading} for rural communities.

Previous studies examining rural mobility tend to distinguish between urban and rural areas using administrative boundaries, built environment contiguity, and population density \cite{chen2022contrasting}. 
While these approaches provide a static perspective on cities and rural areas, in terms of the services they offer or their residential landscapes, they often overlook the interdependencies between rural and urban regions. 
Essential work is needed to infer urban and rural areas using mobility patterns, in order to provide a more coherent framework that can be applied to all countries regardless of their form of urbanisation.

Recent works generally compare accessibility or migration patterns between urban and rural areas, leaving a significant gap in understanding the emergence of inequalities in smaller-scale mobility (i.e. routine travel rather than residential migration).
Only few approaches explore mobility patterns across both urban and rural areas, due their fundamentally distinct characteristics \cite{selod2021rural,busso2021rural}. This poses an additional challenge: comparing rural and urban mobility routing patterns while also considering their inherently distinct topological and demographic features.

We address these limitations by investigating mobility inefficiency inequalities in urban and rural areas, across Colombia, India, and Mexico. To define urban and rural boundaries, we apply network percolation to a country's mobility network, which is defined using origin-destination matrices built from Location Based Data (LBS)\cite{zhang2024netmob2024}. In doing so, we derive an rural-urban classification that is based on how people move through space, rather than solely relying on static characteristics such as population density \cite{chen2022contrasting}. 
This approach enables us to create a universal definition of spatial regions -- from megacities to smaller urban settlements and rural areas -- through the lens of mobility patterns.

Building on this definition, we examine  mobility inefficiency between rural and urban areas, which refers to trips that exceed the optimal shortest path length between their origins and destinations. To accurately characterise inefficiency inequality, we analyse how infrastructure alignment with mobility demands varies across urban and rural contexts. This comprehensive assessment reveals distinct patterns in how transportation infrastructure serves urban and rural needs.
By analysing the angular difference between transit access and urban mobility efficiency vector fields, we pinpoint specific cities and times when access to public transit does not align with mobility efficiency.
In rural settings, that tend to have significantly less access to transit \cite{camarero2019thinking}, we identify inequalities in rural-urban connectivity, providing insights into the specific directions around a city where rural areas face poor access to urban centres and, as a result, limited access to essential resources, such as jobs and hospitals \cite{yan2022spatiotemporal,bastiaanssen2020does}.

Mobility inefficiency serves as a versatile metric, enabling us to not only compare urban-rural disparities but also gain deeper insights into its relationship with urban transit infrastructure and rural accessibility.
Using aggregated, GDPR-compliant mobility data from three low- and middle-income countries (LMIC), we analyse these inequalities in the context of urban versus rural settings.
Our findings reveal that, compared to urban areas, rural areas generally exhibit more inefficiency than their urban counterparts, particularly for longer trips and later in the day. 
These results underscore two critical points: first they emphasise the need to improve urban accessibility for rural areas. 
Second, they highlight the importance of contextually evaluating mobility inefficiency, rather than applying a uniform approach that overlooks the distinct characteristics between urban and rural settings. We reveal four distinct patterns regarding how urban mobility efficiency aligns with public transit accessibility, showing that the majority of cities experience higher transit-efficiency misalignment during working hours on weekdays. 
Finally, we focus on rural areas, highlighting how their proximity and orientation to their closest city, as well as their population can help provide insights as to where natural barriers or urban sprawl may be impacted by rural mobility inefficiency. 
Our framework and findings represent a crucial step in order to ultimately understand the consequences of compounded inequalities \cite{lucas2012transport} (i.e individuals experiencing inequalities in both economic and mobility dimensions). Our results can be further explored \href{https://www.riccardodiclemente.com/projects/rural_urban.html}{here}.

\section*{Results}\label{sec:results}

Existing methods struggle to distinguish urban and rural areas in a manner that can account for diverse urbanisation landscapes. 
This challenge is heightened by the need to  understand how mobility features are expressed in urban and rural settings in order to appropriately assess mobility patterns between areas.
We leverage origin-destination (OD) mobility data, provided by Cuebiq \cite{zhang2024netmob2024} for two purposes (1) to create a mobility network to define urban and rural boundaries and (2) to compare mobility patterns between these two areas.
The data provided exists at a spatial scale for geohash 5 (approx. 25$km^2$ squares) and at 3-hour time intervals, covering November and December 2019. This particular dataset is useful in that its spatio-temporal resolution is coarse enough to be more privacy preserving than individual-level trajectories, but detailed enough to provide insights about rural-urban inequalities.
We utilise transit stop data, obtained from OpenStreetMap \cite{OpenStreetMap}, to calculate an estimate of transit accessibility, which is used to investigate how well transit infrastructure aligns with efficient mobility routing.
To define urban and rural boundaries, we apply network percolation to each country's mobility network, revealing distinct urban regions (see \hyperref[ssec:methods_percolation]{Methods} for details, and \ref{ssec:defining_urban_rural} for urban-rural boundaries in India and Mexico). 
The accuracy of each country's rural-urban classification is validated in the Supplementary Materials, where we compare our approach to the GURS dataset \cite{liu2024global}, which utilizes established data sources like nighttime satellite imagery and built-up surfaces (see \ref{ssec:validating_classification} for details). We assess the provided mobility data for biases by comparing mobility volume to population in each geohash (see \ref{ssec:mobility_bias} for details), which highlights an overall positive correlation between the two dimensions. Given our definition of urban and rural areas, we observe similar trip-length distributions between the two types of areas, but much a lower trip volume for rural areas (Section \ref{sec:mob_ineff_chars}).

\subsection*{Mobility Inefficiency in Rural and Urban Areas}

Understanding how individuals travel from urban and rural areas is crucial.
However, inadequate connections between these regions can significantly increase travel distances for rural residents, while congested urban infrastructure or poor public transit options can create inequitable access to opportunities within cities. 
To accurately assess these disparities, we focus on analysing inequalities in trip lengths rather than simply measuring differences in time spent travelling \cite{niedzielski2014travel}.
This approach avoids assumptions about individual travel modes and addresses the spatial context of mobility. 
We introduce "mobility inefficiency," which highlights the discrepancy between ideally efficient travel and the reality of human mobility patterns that can be impacted by public transportation, congestion, limited infrastructure, or travel preferences. 
To estimate mobility inefficiency for a given OD pair, we leverage stochastic methods to calculate the average optimal shortest path length, comparing it to empirical trip lengths. By aggregating mobility inefficiency for all trips starting from a given geohash, we can also determine the average inefficiency level for that origin, weighting inefficiencies based on the volume of trips travelling to each destination (see \hyperref[ssec:methods_inefficiency]{Methods} for details). 
Mobility inefficiency ($\phi$) is unbounded, with larger, positive values indicating more inefficient trips and negative ones reflecting efficient trips from proximate origin and destination points.

\begin{figure}
    \centering
    \includegraphics[width=0.9\textwidth]{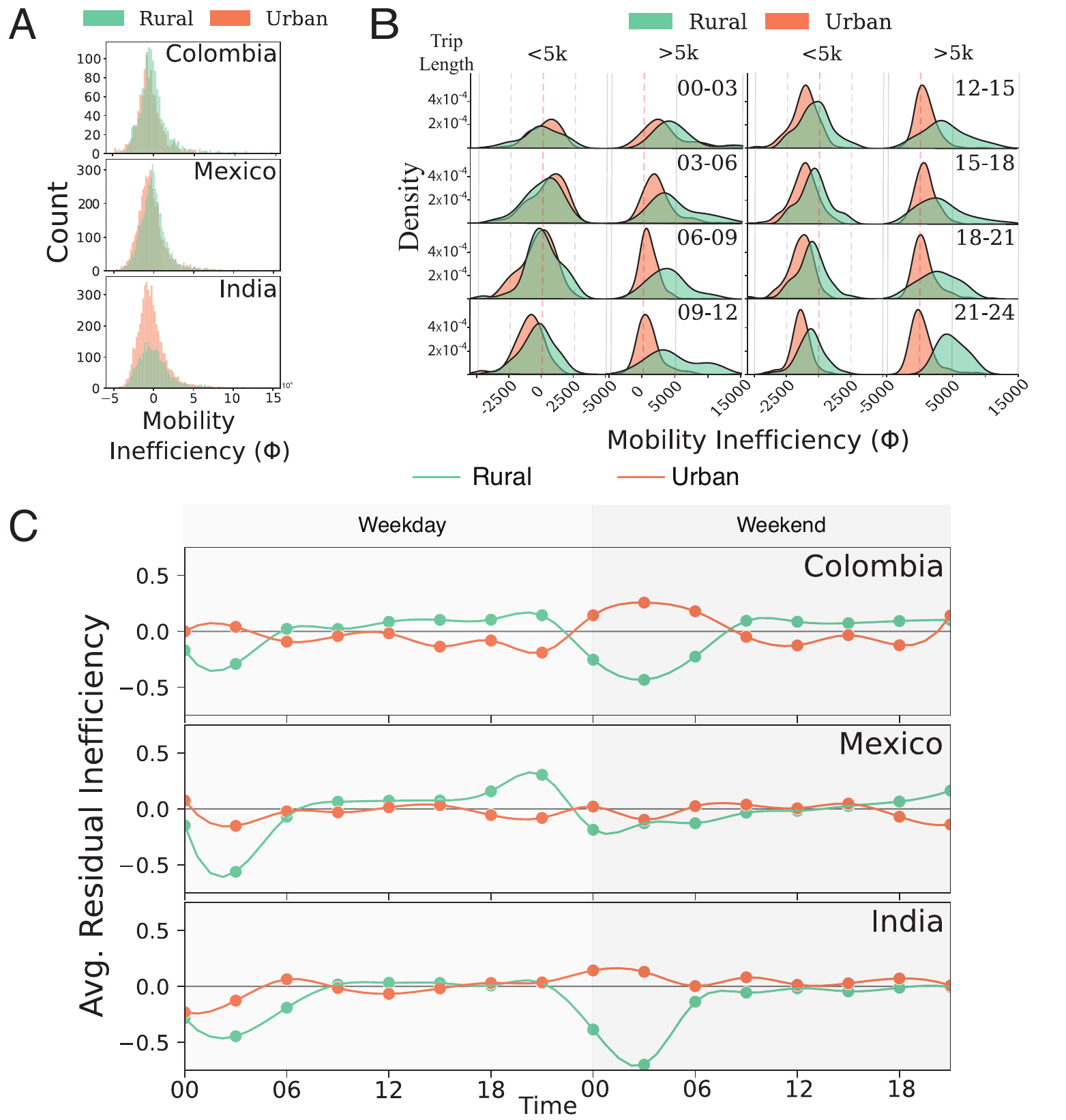}
    \caption{\textbf{Rural-urban inequalities are more apparent for longer trips, later in the day} \textbf{A)} Disparities in mobility inefficiency remain hidden in country-level analyses that do not account for time-based disaggregation. \textbf{B)} Rural areas experience higher inefficiency later in the day. \textbf{C)} Residual inefficiency highlights country-level patterns between urban and rural areas, when normalising for spatio-temporal features of inefficiency.}
    \label{fig:distance_inefficiency}
\end{figure}

We do not find any apparent disparities in inefficiency distributions across the two areas, shown in Figure \ref{fig:distance_inefficiency}A (the statistical properties of these distributions are describes in Section \ref{sec:mob_ineff_chars} of the Supplementary Materials).
To understand whether this is an artefact of aggregating mobility characteristics of trips occurring at different hours of the day, we investigate rural-urban inefficiency features at different times and trip lengths.
Rural areas consistently show higher mobility inefficiency, particularly for trips over 5km, indicating a potential gap in infrastructure for longer trips from rural regions.  Section \ref{ssec:distanct_time_dependent_inefficiency} in the Supplementary Materials highlights how these findings are not present when considering each of the feature (time and trip length) separately and only emerge when mobility inefficiency is studied along both dimensions. As shown in Figure \ref{fig:infrastructure_inefficiency}B, this inefficiency is more pronounced throughout the day compared to urban areas, on an average weekday in Colombia. 
Short trips under 5km in rural areas begin to display inefficiencies starting at 6 AM. These findings are consistent for rural and urban areas in India and Mexico (see Figure \ref{fig:si_ur_mobIneff_temporalSpatial} in the SM for details).

To quantify the rural-urban disparity in mobility inefficiency we define the residual inefficiency, which identifies deviations from typical mobility behaviour, accounting for general mobility behaviour in a focal area and during a given time period. 
Residual inefficiency ($\phi_{g}^{norm}(t)$) of a geohash, $g$, at time, $t$ is computed through two-stage z-score normalisation, first normalising mobility inefficiency with respect to its spatial distribution, then adjusting for time-specific averages.
In doing so, we isolate any temporal and location-based fluctuations (see \hyperref[sssec:methods_residual]{Methods} for Details), a common practice when understanding mobility activity \cite{toole2012inferring}.

Rural communities face heightened mobility inefficiency during commuting hours, particularly between 9AM and 9PM on weekdays in Colombia and Mexico (\ref{fig:distance_inefficiency}C). 
In India, this period of heightened inefficiency is more limited, lasting only from 9AM to 3PM. This shorter time frame in India could be due to a combination of factors. For instance, in India, lower economic activity in the evenings, identified through satellite imagery, can reduce transport demand \cite{alder2017effect}. Furthermore, transportation options in rural India are scarce after 4PM \cite{maretic2020integrated}. Additionally, the agricultural nature of much of  India's rural workforce might lead to earlier curtailment of mobility due to high afternoon temperatures \cite{aayog2017changing}. In contrast, rural areas in the Latin American countries tend to have more mobility later in the day, shown through later transit service in Mexico extending to 7PM and later business hours in Colombia \cite{oecd_colombia}.
Despite the differences between countries, our finding suggests that, generally, rural areas could benefit from improved connectivity to employment locations in order to reduce this gap in mobility inefficiency.

\subsection*{Urban Mobility Infrastructure}
Revealing the difference between rural and urban inefficiencies, with respect to trip length, is only a starting point as it does not provide enough insight into the specific infrastructure accessibility that may contribute to higher levels of mobility inefficiency.
To delve deeper into the relationship between infrastructure access, which in our case in public transportation, and mobility inefficiency we shift our focus to urban areas. 
To explore this concept, we need to measure the level of transit infrastructure that is accessible within an urban area.

\begin{figure}
    \centering
    \includegraphics[width=0.95\textwidth]{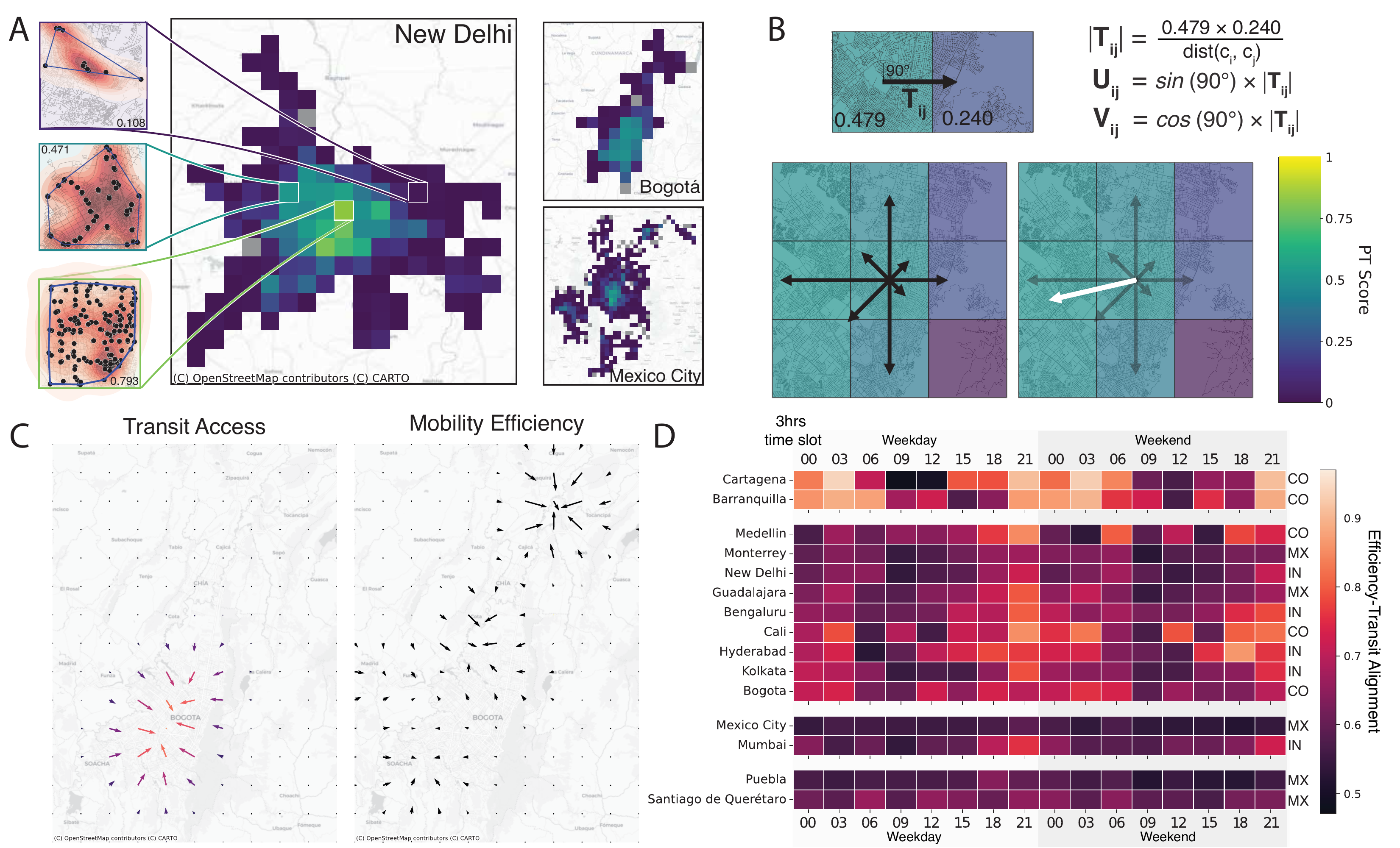}
    \caption{\textbf{Transit access alignment with mobility efficiency tends to be lower during work commuting hours on weekdays.} \textbf{A)} Heat map of transit infrastructure access score in New Delhi, Bogotá, and Mexico City. Left most panels include example of geohashes that with low, medium, and high access scores \textbf{B)} Process of defining a geohash's transit or efficiency vector as defined by Equation \ref{eq:vf} \textbf{C)} Transit access and mobility efficiency vector fields ($\textbf{F}^{PT}$ and $\textbf{F}^{-\phi}_{t}$, respectively)  in Bogotá, Colombia at 9AM on a weekday. \textbf{D)} Alignment of mobility efficiency and transit access across 5 cities in Mexico, India, and Colombia, for different times during weekdays and weekend.}
    \label{fig:infrastructure_inefficiency}
\end{figure}

We introduce the public transit (PT) score, which estimates transit accessibility as the average between the coverage area and density of transit infrastructure in each geohash (see \hyperref[ssec:methods_transit]{Methods} for details). This metric ranges from 0 to 1, with larger values reflecting more access to transit. In this sense, the PT score can capture system accessibility by measuring access to transit stops. 
The coverage area is the ratio of the convex hull area and the geohash area, where the convex hull includes all transit infrastructure in a geohash (left panels of Fig \ref{fig:infrastructure_inefficiency}A).
The transit stop density is defined as the ratio between the number of transit stops and the geohash's area. Section \ref{sec:transit_infrastructre} in the Supplementary Materials distinguishes the difference between the PT Score in rural and urban areas, for all three countries, affirming what is well established: urban areas tend to have more transit access than rural areas. 

The left-most panels of Figure \ref{fig:infrastructure_inefficiency}A discern geohashes in the perimeter of New Delhi to provide a better picture of what a low, medium, and high PT score respectively entails. 
Meanwhile, the geovisualisations in the right-most panels of Figure \ref{fig:infrastructure_inefficiency}A show the spatial distribution of infrastructure scores for New Delhi (India), Bogotá (Colombia), and Mexico City (Mexico), with the centres of cities exhibiting a higher degree of transit access, in terms of coverage and density of transit infrastructure.

At this point, we have a metric for mobility inefficiency ($\phi(t)$) and the public transit ($PT$) access for any given geohash. 
To portray the misalignment between transit access and mobility efficiency in different cities, we can compare the spatial distribution of the two metrics in each urban area, at each time interval.
We construct transit access ($\textbf{F}^{PT}$) and mobility efficiency  ($\textbf{F}^{-\phi}_t$) vector fields using the PT score and the negated mobility inefficiency metric, respectively. 
It is import to note that the mobility efficiency vector field is time-dependent while the transit access one remains static.
We calculate the force between a geohash, $g$, and each of its eight neighbours, representing the interaction strength of adjacent geohashes based on either PT score or mobility inefficiency.
The magnitude of this force is denoted as $T^{PT}_{g}$ and $T^{-\phi}_{g,t}$, respectively. By summing the components of all neighbourhood interactions for a geohash, $g$, we can calculate its resultant vector (Figure \ref{fig:infrastructure_inefficiency}B).
For each city, we compute the resultant transit access vector and efficiency vector, for all geohashes in a city, to obtain a final vector field for transit access ($\textbf{F}^{PT}$) and mobility efficiency ($\textbf{F}^{-\phi}_t$). 
Figure \ref{fig:infrastructure_inefficiency}C shows an example of the transit vector field in Bogotá and the corresponding efficiency vector field at 9AM on a weekday.

We compare transit access and efficiency by computing the average angular difference between each of the geohashes in an urban area (see \hyperref[ssec:methods_transitAlignment]{Methods} for details). Figure \ref{fig:infrastructure_inefficiency}D reveals the alignment between these two features, where larger, brighter values indicate similarity between the two features. To understand which cities are similar in terms of how their efficiency-transit alignment changes throughout the day, we apply spectral clustering to each city's alignment vector for the 48hr period of average weekday and weekend. We identify four main patterns of alignment between transit access and mobility efficiency, displayed from top to bottom of Figure \ref{fig:infrastructure_inefficiency}D. We duplicate this experiment, but implement a naive approach to creating vector fields that are simply based on a focal geohash's local maxima, finding high correlations between both approaches (see Section \ref{sec:vector_field_alt} for details).

The first pattern we discover is cities with high alignment, but less alignment during work hours during regular weekday working hours and from 9AM to 9PM on weekends (Cartagena, Barranquilla). The majority of cities exhibit a pattern of mild alignment: expressing moderate alignment throughout the day and spikes of more alignment from 6PM onwards, on both weekdays and weekends. Cities in Colombia generally show greater alignment between transit access and mobility efficiency than those in other countries. This likely stems from each of the 5 Colombian cities hosting robust bus rapid transit (BRT) systems, which are characterised by high capacities and dedicated bus lanes \cite{muller2014bus}.

The last two trends that emerge are (1) low, orthogonal alignment between transit access and mobility efficiency throughout the entire weekday and weekends (Puebla and Querétaro) and (2) low alignment with moderate alignment with a brief increase in alignment at the end of the day (Mexico City and Mumbai). Puebla and Querétaro display a consistent orthogonal alignment between transit access and mobility efficiency throughout the week. This pattern might stem from several factors: limited transit frequency, as documented in Querétaro \cite{obregon2016impact}, or high reliance on transit that isn't adequately met by current service levels, a common challenge in larger Mexican cities \cite{guerra2018urban}. Job decentralisation could also play a role, contributing to poor job accessibility for transit-dependent low-income workers, as observed in Mexico City \cite{bautista2020commuting}.

\subsection*{Rural Inefficiency in the Urban Periphery}
Rural areas are often overlooked in human mobility studies, underscoring the need for a quantitative approach to understand which types of rural areas experience particularly high levels of inequality, such as mobility inefficiency.
Since rural regions exhibit much lower rates of transit access coverage and mobility data (see Section \ref{sec:transit_infrastructre} in the SM for details), the vector field approach used in urban settings would not be relevant in the rural context.
We focus on rural areas that surround three major cities: Querétaro (MX) New Delhi (IN) and Bogotá (CO) to gauge how natural barriers influence where rural areas can emerge and how well connected they are to nearby urban areas. 
In doing so, we aim to highlight the highly contextual nature of improving rural-urban connectivity, as factors such as city size, sprawl, natural barriers influence where rural areas can emerge and how well connected they are to nearby urban areas. 

To link rural areas to their nearest urban centres, we calculate the distance from each rural geohash to each city's centroid, in a given country.
A city's centroid is determined by averaging the centroids of its geohash cells, weighted by area. For each rural geohash, we determine the city to which it is closest. 
Rural areas are grouped based on their distance from and orientation around the central point of a city. Finally, we compute the average inefficiency for each orientation-distance pair, with concentric circles reflecting different exponential distance bins (Fig \ref{fig:country_comp}A). 
Additionally, surrounding grey bars indicate the cumulative population of rural areas in each direction, emphasising the extent to which city sprawl in various directions affects mobility inefficiency.

\begin{table}
\caption{Regression results for New Delhi, Santiago de Querétaro, and Bogotá, when applying Ordinary Least Squares on spatial features of their surrounding rural areas to predict their respective mobility inefficiency.}
\label{tab:rural_regression}
\begin{center}
\begin{tabular}{llll}
\hline
                     & New Delhi  & Santiago de Querétaro & Bogotá       \\
\hline
const                & -1438.47   & 180.82                & \textbf{-3416.18***}  \\
                     & (1044.01)  & (1345.72)             & (1198.63)    \\
Distance from Centre             & -1.29      & -11.46                & \textbf{21.63**}      \\
                     & (6.81)     & (15.47)               & (8.20)       \\
Population           & \textbf{0.03***}    & 0.04                  & 0.01         \\
                     & (0.01)     & (0.03)                & (0.01)       \\
East-West Orientation      & \textbf{-1662.90**} & \textbf{1972.94*}              & 254.19       \\
                     & (730.53)   & (1083.15)             & (427.20)     \\
North-South Orientation     & -542.27    & 343.54                & \textbf{-1031.48**}   \\
                     & (456.48)   & (693.42)              & (506.08)     \\
Proximity to cities & \textbf{-31.05**}   & -36.44                & \textbf{30.64*}       \\
                     & (13.93)    & (24.44)               & (15.49)      \\
R-squared            & 0.30       & 0.26                  & 0.17         \\

\hline
Standard errors in parentheses. \\
* p<.1, ** p<.05, ***p<.01 \\
\hline
\end{tabular}
\end{center}
\end{table}

Population is critical to consider because it allows us to assess whether inefficiency impacts a large proportion of the surrounding rural population or mainly has an effect or areas with fewer people, potentially indicating low demand for improving mobility infrastructure. Querétaro is a strong example for why population is an important feature, as it only has one rural region to its northwest, which consists of the largest rural population in Querétaro.

Our findings reveal spatial disparities in mobility inefficiency that are unique to each city by using Ordinary Least Squares (OLS) regression to analyse the relationship between various features and inefficiency. 
These features include distance from each city's centroid, population of the rural area, orientation, and its proximity to cities. 
The orientation is decomposed into it's north-south and east-west distance, where positive values indicate a relationship between inefficiency and northern and eastern rural areas, respectively.
The proximity to cities represents the absolute difference in distance between a rural area and its two nearest cities. It quantifies how much closer or farther the rural area is to one city compared to the other. This difference could provide insight into the area's relative isolation or proximity to urban hubs, which might affect its access to resources, services, or infrastructure. We apply to OLS to each of the 15 cities analysed in the previous section, the results for which are presented in Section \ref{sec:ols_extra} of the Supplementary Materials.

Here, we focus on one major city in each of the countries, to underscore how inequalities can be interpreted visually and quantitatively The regression results, in Table \ref{tab:rural_regression}, reveal key differences in how inefficiency is influenced across New Delhi, Santiago de Querétaro, and Bogotá. Distance plays a role in inefficiency, particularly for Bogotá, where being farther away is linked to higher inefficiency.
Meanwhile, population size has a clear impact in New Delhi, suggesting that as the population grows, inefficiency increases, while in the other two cities, this relationship is weaker or not significant.

Directional factors also play a critical role in inefficiency. In New Delhi, east-west positioning significantly affects inefficiency, with areas further west experiencing higher inefficiency, as indicated by the negative coefficient in Table \ref{tab:rural_regression} and the rightmost panel in Figure \ref{fig:country_comp}A.
We observe a negative coefficient for the East-West orientation in Santiago de Querétaro, consistent with the central panel of Figure \ref{fig:country_comp}A, which shows that rural areas southeast of Querétaro, within 32 kilometres of the city centre, experience notably higher inefficiency. This aligns with the topology of Querétaro, which consists of rugged topography in its northeast and southeast areas \cite{juarez2024sustainable}. 
In contrast, rural populations to the southwest and north of the city tend to be smaller and show lower inefficiency. In Bogotá, inefficiency is strongly linked to north-south positioning, which can be linked to the spatial distribution of poverty, with the more affluent population concentrating in the north \cite{guzman2017urban}. We observe few instances of rural areas in the east and south, which can be attributed to the mountainous regions surrounding Bogotá in these regions \cite{andrade2013assembling}. Additionally, the relative difference in distances within the city is an important factor in both New Delhi and Bogotá, further shaping inefficiency patterns. 

In New Delhi, a negative coefficient suggests that rural areas with a greater distance disparity between the two cities tend to experience lower inefficiency, possibly because they are less reliant on resources or services from urban centres.
In contrast, Bogotá’s positive coefficient indicates that areas with a larger distance gap experience higher inefficiency, likely due to a greater reliance on one city for resources, leading to challenges in accessing services and infrastructure. 
These contrasting effects highlight how the availability of resources from urban areas shapes inefficiency in different contexts. While the model captures some variation in inefficiency, other unaccounted factors, such as topological features, likely contribute to inefficiency in all three cities.

\subsection*{Weekday-Weekend Shifts in Mobility Inefficiency}
Thus far, we have compared urban and rural inefficiency across 3-hour time intervals. 
Here, we examine how inefficiency is influenced by broader societal rhythms, comparing inefficiency between weekdays and weekends to identify patterns of mobility and resource usage that may vary due to differences in work-related activities, leisure behaviours, and transportation demand.
By using the discretised version of $\phi$ and the similarity score $\text{sim}(\vec{\Phi_1}, \vec{\Phi_2})$, we assess how mobility inefficiency changes between weekdays and weekends, in cities and countries. 

\begin{figure}
    \centering
    \includegraphics[width=0.9\linewidth]{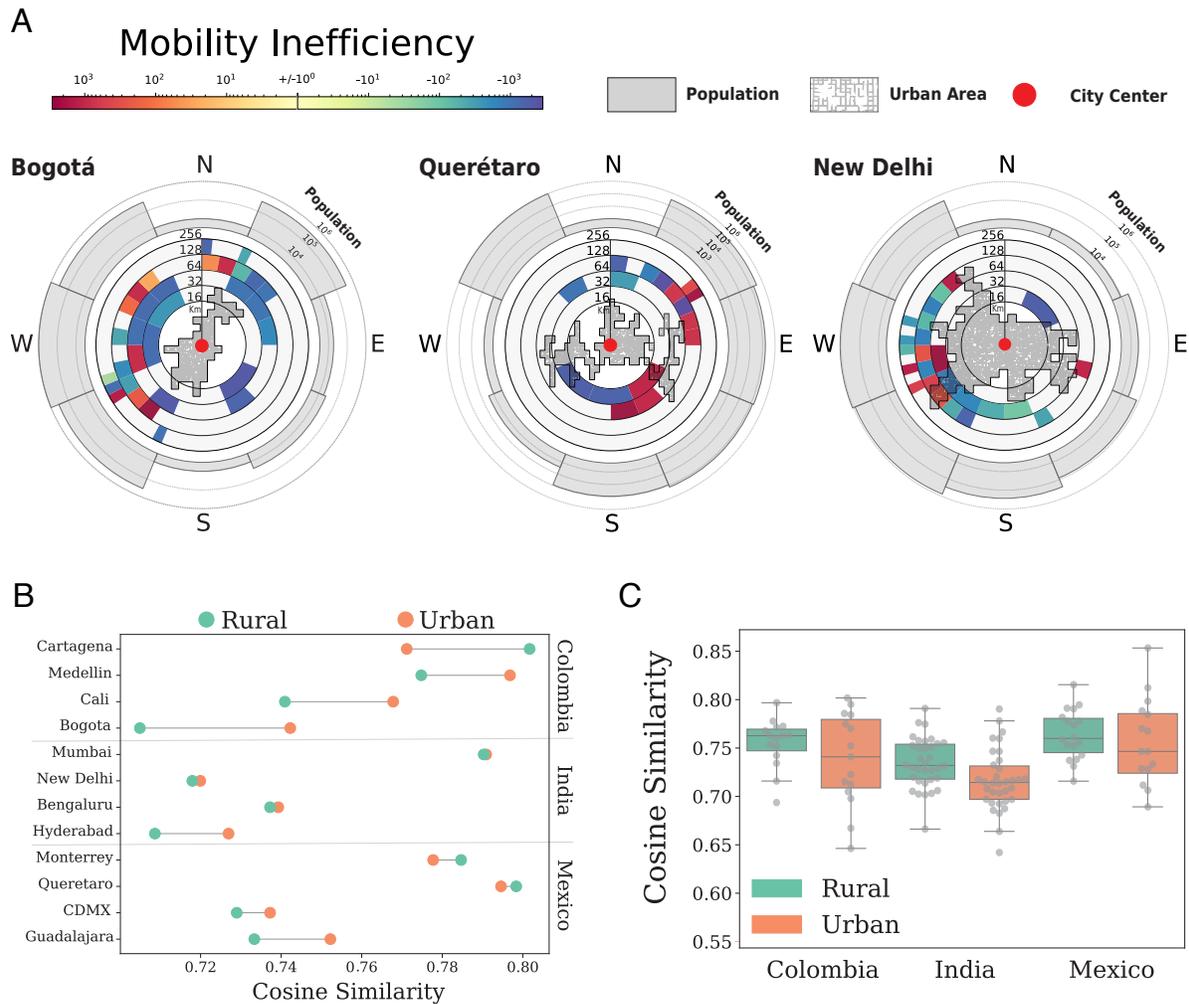}
    \caption{\textbf{Focusing on rural mobility inefficiency highlights disparities across rural areas while comparing changes between weekday and weekend mobility inefficiency at the city and country levels exposes potential rural-urban dependencies.} \textbf{A)} Mobility inefficiency in rural areas within a 256km buffer of \textbf{B)} City-level differences in how urban and rural mobility inefficiencies shift from weekend to weekday. \textbf{C)} Country-level comparisons show less pronounced weekend-weekday differences in inefficiency between urban and rural areas.}
    \label{fig:country_comp}
\end{figure}

Figure \ref{fig:country_comp}B shows how similar a city is during weekdays to weekends for 12 major cities (orange point) and their relative rural areas (green points) across India, Colombia, and Mexico. 
We found that, in Colombian cities, urban and nearby rural areas show notable differences in consistency of mobility inefficiency, with rural areas experiencing greater changes in inefficiency between weekdays and weekends. This indicates that urban areas tend to have more consistent inefficiency patterns, while rural areas may need more flexible mobility solutions that account for varying trip purposes, such as weekend tourism.
In Colombian cities, mobility inefficiency is more consistent in urban areas, while rural areas see greater variation between weekdays and weekends. This suggests that rural areas may need more flexible mobility solutions to accommodate changing trip purposes, such as a decrease in the volume of commutes and an more leisurely trips during the weekends \cite{toger2023inequality}.

An exception is Cartagena, where the city centre is more stable than the surrounding suburban and rural areas.
Meanwhile, cities in India, barring Hyderabad, experience more stable temporal fluctuations in mobility inefficiency compared to their rural surroundings. This could be an artefact of rural reliance on urban centres, due to their concentration of essential services \cite{shaban2020india}, leading to stable mobility inefficiency in rural areas since travel patterns remain consistent from weekday to weekend. 
In Mexican cities, the difference in inefficiency patterns between weekdays and weekends is relatively similar for urban and rural areas.

Interestingly enough, while intricate differences emerge when looking at results at city-scale levels (Figure \ref{fig:country_comp}B), when we compute the average cosine similarity at a country level (Figure \ref{fig:country_comp}C), considering all the urban areas in a country, we find trends across the three countries. Urban areas consistently exhibit smaller cosine similarity, indicating larger fluctuations in inefficiency between weekdays and weekends. This robust pattern may reflect commuting behaviours, where urban areas experience more pronounced inefficiency changes due to shifts in daily travel patterns and higher reliance on weekday commutes, while rural areas show more stable inefficiency across the week.

\section*{Discussion}

Understanding the differences in urban and rural mobility patterns requires analysing these areas not only in terms of their interdependence, but also within their respective contexts. 
Studies examining both urban and rural areas focus primarily on internal migration or accessibility \cite{selod2021rural,cattaneo2021global,camarero2019thinking}.
A one-size-fits-all approach to accessibility is inadequate due to its subjective nature \cite{pot2024far} (i.e. opinions on transit service and levels of car dependence differ between rural and urban areas). In our work, we expand on this notion, developing a metric for mobility inefficiency, which reflects how empirical trip lengths deviate from the optimal, shortest-path, considering the structure of local road networks. With mobility inefficiency, we can evaluate how inefficiencies in rural areas are more pronounced than their urban counterparts, for longer trips, during the later hours of the day. These clear differences in mobility inefficiency could be an artefact of the same mobility metric carrying different meanings in the two types of regions due their contrasting needs \cite{gimenez2024daily} and resources 
\cite{cattaneo2021global,vitale2020improving,moseley2023accessibility}. Furthermore, our work uses mobility inefficiency as a tool to quantify the lack of necessary infrastructure in rural regions for supporting efficient long-distance travel \cite{camarero2019thinking,bauchinger2021developing}, highlighting higher levels of inefficiency in rural areas for long trips.

The key contribution of this work lies in interpreting mobility inefficiency inequalities in the context of their rural or urban environment. We determine four distinct patterns, in terms of how aligned public transit access is with mobility efficiency, using the average angular difference between their respective vector fields.
We reveal that cities in Colombia tend to be more aligned, which could be a reflection of its cities adopting bus rapid transit systems \cite{muller2014bus}. 
Meanwhile, 3 of the 5 Mexican cities have orthogonal alignments for the majority of the weekday and weekends, which could be an artefact of poor transit service or decentralised employment opportunities \cite{bautista2020commuting}.

Meanwhile, for rural areas we apply a local analysis on each city's periphery to explore how rural inefficiency relates its orientation, distance and population, revealing that rural inefficiency, for areas surrounding New Delhi and Querétaro, are not associated with distance from the respective city centres. Rural areas surrounding Bogotá and Querétaro that have high mobility inefficiency do not exhibit any strong link based on their population. By applying OLS to the rural areas surrounding New Delhi, Querétaro and Bogotá, we find the inefficiency increases as distance from the city centre grows in a westerly, easterly, and southerly direction, respectively. This suggests that factors beyond distance and population density influence rural mobility inefficiency, potentially including land use patterns \cite{juarez2024sustainable}, natural barriers \cite{ andrade2013assembling}, and socioeconomic disparities \cite{guzman2017urban}. Further investigation is needed to fully understand these complex relationships.

To better understand how broader patterns of mobility inefficiency may arise, we focus on the variation in mobility efficiency between weekdays and weekends, finding that urban areas generally exhibit more consistent inefficiency patterns over time. In contrast, rural areas showed greater fluctuations, suggesting that rural mobility may be more sensitive to variations in trip purpose, such as increased travel for leisure or tourism on weekends \cite{gimenez2024daily}. 
This finding could inform more targeted rural mobility interventions that cater to the changing needs of rural populations during different times of the week.
Interestingly, while we observed significant disparities in mobility inefficiency at the city scale, country-level analyses revealed more uniform patterns of inefficiency across Colombia, India, and Mexico. 
This highlights the importance of conducting mobility studies at more localised scales to capture the specific infrastructural and mobility challenges faced by individual cities and regions.

Our research introduces a novel perspective on understanding mobility inequalities by examining both urban and rural areas using a consistent metric. This approach underscores the critical need for contextually analysing urban-rural divides, shedding light on how these disparities manifest across different geographic scales.
However, our findings also raise important questions for future research. 
First, while the mobility inefficiency metric provides valuable insights, it assumes that all trips should ideally follow the shortest path.
Future work could explore alternative models of inefficiency that take into account variations in travel purposes and modes of transportation. 
Additionally, while our study highlights infrastructure as a key determinant of mobility efficiency, further research is needed to unpack the specific types of infrastructure—such as road quality, transit service, or multimodal networks—that contribute most to reducing inefficiencies.

\section*{Methods}\label{sec:methods}

\subsection*{Dataset}\label{ssec:methods_data}

To define urban and rural areas with respect to mobility patterns as well as explore differences in trip behaviour between these environments, we require country-wide mobility data. Aggregated data was provided by
Cuebiq Social Impact as part of the NetMob 2024 conference \cite{zhang2024netmob2024}. Data is collected with the informed
consent of anonymous users who have opted-in to anonymized data collection for research purposes. 
In order to further preserve the privacy of users, all data has been aggregated by the data provider to the geohash 5 or H3 7 level and does not include any individual-level data records. We use the origin-destination data (November 1, 2019 to December 31, 2019), provided by Cuebiq, at 3-hour intervals and a spatial scale of geohash 5 (4.89$\text{km}^2$ squares). 
Each row of the provided data details the origin geohash, destination geohash, trip count, and statistical properties (mean, median, standard deviation) of the trip lengths and trip times. Table \ref{tab:trip_chars} highlights the distribution of trips within the same geohash (column 3), to neighbouring geohashes (column 4), and to other geohashes that do not share an edge or vertex with the origin geohash (column 5), showing how the majority of trips in the data start and end in the same geohash. This underscores how the low spatial resolution of geohash 5 may obscure finer inequalities within each cell. However, since we are comparing disparities in mobility patterns between urban and rural areas, high spatial resolution is less critical. 
\begin{table}[]
    \centering
    \begin{tabular}{lllll}
\toprule
Country & No. Trips & \multicolumn{3}{c}{\textbf{Destination Distribution} (\%)}\\
 &  & Within & Adjacent & Other \\
\midrule
Colombia & 311,640,800 & 81.7 & 16.58 & 1.73 \\
Mexico & 9,293,634,700 & 82.77 & 16.26 & 0.97 \\
India & 4,273,311,000 & 94.74 & 5.23 & 0.03 \\
\bottomrule
\end{tabular}

    \caption{Percentage of trips to within the same geohash, to neighbouring geohashes, to non-adjacent geohashes.}
    \label{tab:trip_chars}
\end{table}

\begin{figure}
    \centering
    \includegraphics[width=0.5\linewidth]{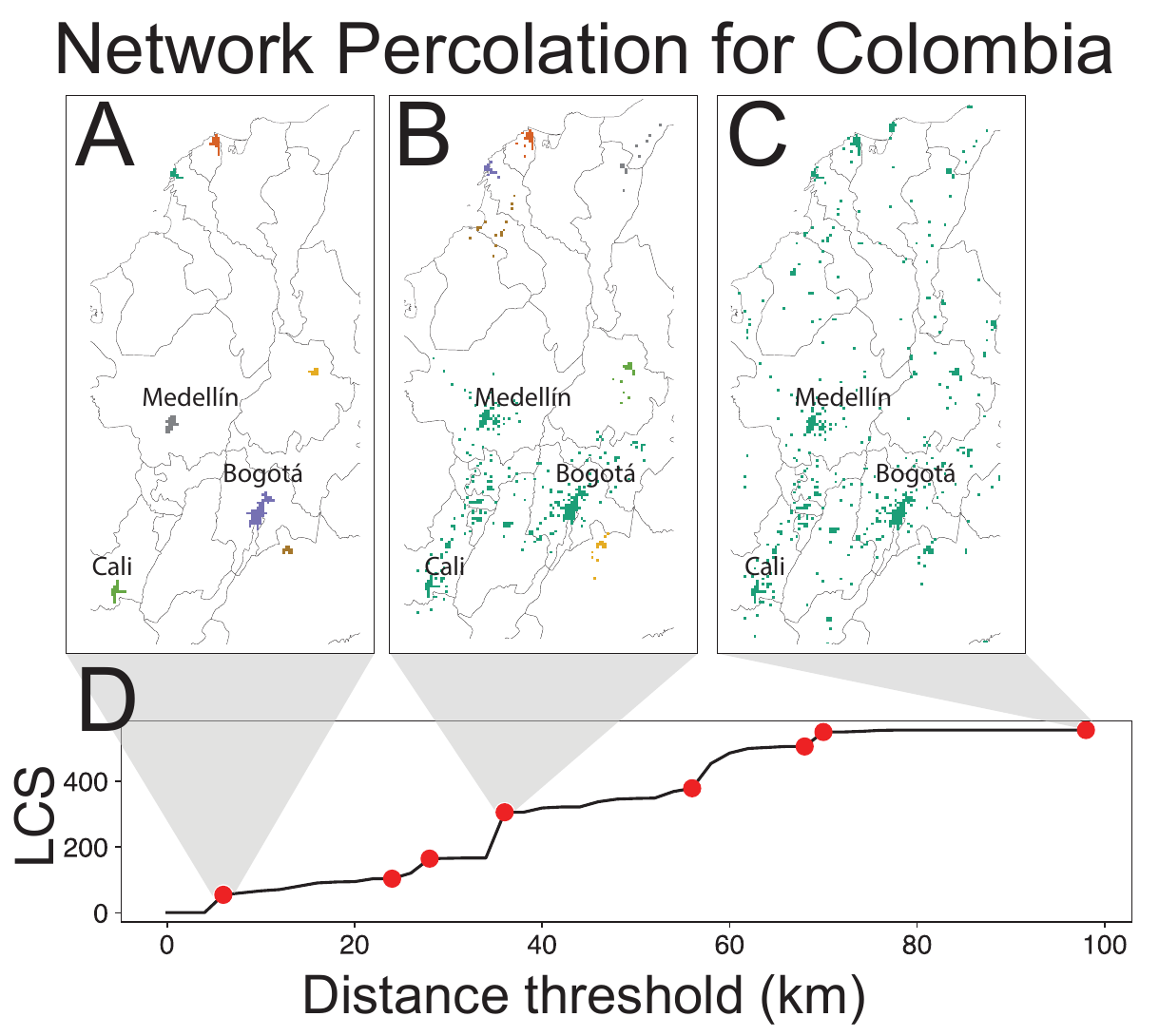}
    \caption{\textbf{Defining urban boundaries in Colombia by applying percolation to mobility networks.} \textbf{A-C)} Urban boundaries in Colombia at the first, fourth, and last critical point in Colombia, respectively. \textbf{D} Largest component size (LCS) at each distance threshold, with red points indicating each critical point in Colombia.}
    \label{fig:percolation}
\end{figure}
\subsection*{Defining Urban and Rural Areas}\label{ssec:methods_percolation}

To identify urban and rural areas from a mobility perspective, we construct a network such that nodes reflect visited geohash cells, provided by the origin-destination data. In doing so, we filter out areas that do not exhibit mobility patterns, which conventional approaches may consider urban due their built-up surface (see Section \ref{ssec:validating_classification} in the Supplementary Materials for details). Network edges, weighted by distance (km), connect neighbouring geohashes and link the closest geohashes between disconnected components, yielding a final network encompassing mobility and geospatial characteristics. We apply network percolation \cite{arcaute2016cities} on this network at different distance thresholds, revealing the hierarchical structure of urban areas (Figures \ref{fig:percolation}A-C). As the threshold increases, clusters evolve from distinct cities to larger regions, eventually encompassing all geohashes in Colombia at the 100km threshold. Percolation theory typically deals with infinite systems in which a critical point can be discovered when a giant cluster forms \cite{stauffer2018introduction}. However, when applying percolation to urban networks, which are finite in nature, various critical points can emerge (Figure \ref{fig:percolation}D), identifying distinct hierarchical levels. In this manner, we define urban areas based on the clusters that form at the first critical point for each country (Figures \ref{fig:percolation}A-C). Specifically, the first critical point occurs at a 6km threshold for Colombia and India, and at 8km for Mexico (see Supplementary Materials for critical points of all countries). Geohash cells not included in the major urban clusters are classified as rural, allowing us to compare mobility patterns in rural and urban areas by associating each rural area to its closest city.

\subsection*{The Inefficiency Score}\label{ssec:methods_inefficiency}
Defining urban and rural boundaries enables us to explore how individuals travelling from these regions may exhibit certain mobility patterns.
Inadequate links to urban areas can exacerbate the distance that individuals from rural areas travel to achieve their mobility demands. Meanwhile, feature such as poor transit infrastructure or high congestion within urban areas may impact 
We focus on analysing inequalities in trip lengths, rather than trip times, to avoid making assumptions about travel. Furthermore, we want to avoid simply measuring differences in trip length between urban and rural areas as doing so may just capture differences in spatial dependencies between the two area types. 
Thus, we propose a metric for "mobility inefficiency", $\phi$, for a given trip at time, $t$, from an origin, $o$ to a destination, $d$:
\begin{equation}
    \phi(o,d,t) = \text{EP}(o,d,t)- \text{SP}(o,d,t),
\end{equation}
which captures the difference between empirical trip lengths $EP$ and an estimated benchmark length $SP$. The benchmark, $SP$, is measured by randomly sampling 100 nodes from the road networks of the origin and destination geohashes in the provided data. This benchmark reflects an ideal trip length, assuming the shortest path is the optimal one and that trips' origins and destinations are uniformly distributed across the respective geohashes' road networks. The mobility inefficiency is theoretically unbounded, with positive values of $\phi$, indicating that empirical trips are longer than their optimal counterparts and are consequently inefficient. It can also be aggregated to the the origin level such that $\phi(o,t)$  reflects the weighted mean inefficiency of all trips originating in geohash $o$ at time $t$, with respect to trips counts between OD pairs.

Ultimately, this formulation of mobility inefficiency allows us to analyse inequalities and different temporal and spatial scales: comparing city and country level inequalities at different times. In doing so, we reveal how city-level comparisons are most informative for identifying cities and their surrounding rural areas that experience the largest change in efficiency.  

\subsubsection*{Residual Inefficiency}\label{sssec:methods_residual}
To evaluate disparities in mobility inefficiency between rural and urban areas, we adopt a methodology inspired by the concept of residual activity \cite{toole2012inferring}. In the context of mobility inefficiency, we extend this concept to examine deviations from travel inefficiencies across different regions and time periods. Residual inefficiency (\(R\)) is computed as the z-score normalised mobility inefficiency ($\phi_{g}^{norm}(t)$) of a geohash, $g$, at time, $t$, with respect to the average normalised inefficiency in the entire area at a given time ($ \phi^{res}(g,t)$):
\begin{subequations}
    \begin{equation}
        \phi^{norm}(g,t) = \frac{\phi(g,t) - \mu_{\phi(g)}}{\sigma_{\phi(g)}}
    \label{eq:res_ineff_1}
    \end{equation}

    \begin{equation}
        \phi^{res}(g,t) = \phi^{norm}(g,t) - \overline{\phi} {}^{norm}(t)
    \label{eq:res_ineff_2}
    \end{equation}
\label{eq:res_ineff}
\end{subequations}

The z-score normalisation in Equation \ref{eq:res_ineff_1} accounts for local variability in inefficiency, ensuring that comparisons are not skewed by inherent differences in mobility patterns across geohashes. Temporal variations in inefficiency trend, such as the increased inefficiency during commuting hours, are accounted for in Equation \ref{eq:res_ineff_2}. This approach highlights exceptionally high or low inefficiency values by adjusting for both spatial and temporal trends. Finally, residual inefficiency for a given area type is computed by averaging all residual inefficiency values for geohashes classified as urban or rural.

\subsubsection*{Discretise the Inefficiency Score}\label{sssec:methods_inefficiencyQuartiles}

In order to be able to compare $\phi$ across cities and countries, we discretized $\phi$ into quartiles. Each $\phi(o,d,t)$ is assigned to a group in $G = \{g_1, g_2, g_3, g_4\}$ with a function of the form $f(\phi(o,d,t)) \rightarrow G$. The split can be done both at the city- or country-scale, depending on the analysis we desire to carry out.  
In both cases, $g_1$, consist of very inefficient areas, in $g_2$ slightly inefficient areas, in $g_3$ slightly efficient areas, and, $g_4$ reflects very efficient zones. With this discretization of $\phi$, we can better compare cities and countries and to measure how efficiency changes in different temporal contexts. 
To this end, we leverage the cosine similarity. Given an area of interest (e.g., cities, countries) and two different contexts $c_1$ and $c_2$ (e.g., weekend and weekdays), we build the vectors $\vec{\Phi_{c_1}}$ and $\vec{\Phi_{c_2}}$, which contain the list of $f(\phi(o,d,t))$ for all the $o$ and $d$ in the area of interest.
The similarity between the $\vec{\Phi_{c_1}}$ and $\vec{\Phi_{c_2}}$ is computed as:

$$\text{sim}(\vec{\Phi_1}, \vec{\Phi_2}) = \frac{\vec{\Phi_1} \cdot \vec{\Phi_2}}{\|\vec{\Phi_1}\| \|\vec{\Phi2}\|}$$

\noindent $\text{sim}(\vec{\Phi_1}, \vec{\Phi_2})$ is bounded in $(0,1)$ as vector components cannot be negative. Values close to $1$ indicate a high similarity between the two vectors, while values $0$ indicate dissimilar vectors.

\subsection*{The Public Transit Score}\label{ssec:methods_transit}
The public transit (PT) score, ranging from 0 to 1, serves as a proxy for transit access, by considering how the density and coverage of transit stops in urban areas. For each urban geohash, we retrieve transit stops from OpenStreetMap \cite{OpenStreetMap}. To estimate the coverage area, we divide the area for the convex hull of the transit stops by the area of the geohash. We define a geohash's transit stop density as the ratio between the number of transit stops and the geohash's area. The density is min-max normalised over all geohashes in a country. The PT score is determined by averaging the coverage area and normalised density of transit stops in a geohash. Figure \ref{fig:si_pt_score_distr} in the Supplementary Materials shows the distribution of the PT score for all urban geohashes in Colombia, Mexico, and India.

\subsection*{Efficiency-Transit Alignment}\label{ssec:methods_transitAlignment}
Efficiency-Transit Alignment (ETA) conceptually reflects how similar mobility inefficiencies and transit access are across all geohashes in an urban area. To achieve this, we create vector fields representing both transit access and efficiency. Using concepts from the gravity model \cite{}, we determine the interaction strength between neighbouring geohashes, with the the interaction reflecting either transit access, $PT$, of two geohashes or their mobility inefficiency, $\phi(o,t)$. Two geohashes are considered neighbours if they are connected through Queen contiguity (i.e., they share a vertex, corner, or edge). In this manner, We can define a vector, $T^{PT}_{ij}=(U_{ij}, V_{ij})$, to reflect the attractive force between two adjacent geohashes $i$ and $j$:

\begin{subequations}
\begin{equation}
   |T^{PT}_{ij}| = \frac{PT_i * PT_j}{\textit{dist}(c_i, c_j)^2}, 
   \label{eq:vf_mag}
\end{equation}

\begin{equation}
   U^{PT}_{ij} = sin(\theta_{ij}) * |T^{PT}_{ij}|
   \label{eq:cos}
\end{equation}

\begin{equation}
   V^{PT}_{ij} = cos(\theta_{ij}) * |T^{PT}_{ij}|
   \label{eq:sin}
\end{equation}

\label{eq:vf}
\end{subequations}

where $dist(c_i, c_j)$  is the distance between the centroids 
$c_i$ and $c_j$ of the two geohashes. Equations \ref{eq:cos} and \ref{eq:sin}, which measure the directional components of $T^{PT}_{ij}$, are defined using the angle between the two geohashes ($\theta_{ij}$). We can then define, $\vv{T}^{PT}_i$, as the resultant vector for all neighbouring geohashes to $i$, where neighbours are defined using Queen contiguity (e.g. any geohash that shares a vertex with $i$). Figures \ref{fig:infrastructure_inefficiency}B-C illustrate this process, with the resultant vector shown in white. The transit access vector field, 
$\textbf{F}^{PT} = (P_T, Q_T)$, is built for all geohashes in an urban area, where $P_T$ and $Q_T$ reflect the horizontal and vertical components of the vector field, respectively.
Similarly, the mobility efficiency vector field, $\textbf{F}^{-\phi}_t = (P_E, Q_E)$, can be defined using Equation \ref{eq:vf}, by incorporating the time interval being considered, $t$, and byreplacing the PT score with mobility efficiency ($-\phi(i,t)$). It should be noted that, unlike the mobility efficiency vector field, the transit access vector field remains constant over time.

To quantitatively compare the angular difference of vector fields constructed based on local public transit access and mobility efficiency we utilise the discrete integral of their angular difference, for a given time interval. Considering that the each geohash roughly reflects 25 $km^2$ cells, the average angular difference, in degrees, over an entire urban area can be described as:

\begin{equation}
    \overline{\text{I}_{\theta}} = \frac{1}{|G|}\sum_g^G \frac{180}{\pi}\cos^{-1}\left(\frac{\textbf{F}^{PT}_{g}\cdot\textbf{F}^{-\phi}_{g,t}}{\norm{\textbf{F}^{PT}_{g}} \norm{\textbf{F}^{-\phi}_{g,t}}}\right),
\end{equation}

\noindent where $G$ reflects the set of geohashes in the urban area. $\overline{\text{I}_{\theta}}$ is naturally bounded between 0 and 180 degrees. To determine ETA, we take the converse of $\overline{\text{I}_{\theta}}$ (since larger angular differences reflect less alignment) and min-max normalise this value with respect to its bounds. 
ETA, then, is bounded between 0 and 1, with values approaching 1 being indicative of smaller angular differences between the transit access and mobility inefficiency vector fields.

\section*{Data availability}

Data was made available from Cuebiq and the World Bank as part of the NetMob 2024 Data Challenge. OpenStreetMap data is openly accessible via \href{https://www.openstreetmap.org/}{https://www.openstreetmap.org/}. Population data from WorldPop is available at \href{https://www.worldpop.org/}{https://www.worldpop.org/}. GURS data on urban and rural settlements are available via \href{https://zenodo.org/records/11160893}{https://zenodo.org/records/11160893}. The code to reproduce this data is available at \href{https://github.com/ComplexConnectionsLab/urban_rural_inefficiency}{https://github.com/ComplexConnectionsLab/urban\_rural\_inefficiency}. The website to explore our results can be accessed via \href{https://www.riccardodiclemente.com/projects/rural_urban.html}{https://www.riccardodiclemente.com/projects/rural\_urban.html}.

\section*{Acknowledgements}

The authors would like to acknowledge Cuebiq, the World Bank and NetMob 2024, through which the data used in this manuscript was made available.

\section*{Author contributions statement}

N.I., M.L. and R.D.C. conceived the experiment(s),  N.I. analysed the results, M.L. analysed the weekday-weekend comparisons, N.I. and R.D.C. visualised the results.  R.D.C. supervised the project. All authors reviewed the manuscript. 

\section*{Ethics Declaration}

\noindent \textbf{Competing interests} 
The authors declare no competing interests.





\include{si}

\bibliography{biblio,si_bib}
\end{document}

%% file: si.tex
\newcommand{\lto}[1]{\longrightarrow#1}
\newcommand{\rto}[1]{\longleftarrow#1}
\newcommand{\imply}[1]{\Longrightarrow#1}
\newcommand{\Eq}[1]{Eq.\,(\ref{#1})}
\newcommand{\Fig}[1]{Fig.\,\ref{#1}}
\newcommand{\tr}[1]{\text{Tr}$#1$}
\setcounter{figure}{0}
\renewcommand{\figurename}{Fig.}
\renewcommand{\thefigure}{S\arabic{figure}}
\renewcommand{\theequation}{S\arabic{equation}}
\renewcommand{\thetable}{S\arabic{table}}

\makeatletter
\renewcommand\@seccntformat[1]{\csname the#1\endcsname \quad}
\makeatother

\renewcommand*{\thesection}{}
\renewcommand*{\thesubsection}{S\arabic{section}.\Alph{subsection}}
\renewcommand*{\thesubsubsection}{S\arabic{section}.\Alph{subsection}.\arabic{subsubsection}}

\newcommand{\heading}[1]{{\vspace{0.25truecm}\noindent\textbf{#1.}}}



\begin{center}
    {\LARGE \textbf{Supplementary Materials: Understanding Urban-Rural Disparities in Mobility Inefficiency for Colombia, Mexico, and India}}\\[1em]
    {\large Nandini Iyer, Massimiliano Luca, Riccardo Di Clemente} \\[5em]
\end{center}

\section{Supplementary Note 1 Rural-Urban Population Characteristics}
\subsection{Defining Urban and Rural Geohashes}\label{ssec:defining_urban_rural}
\begin{figure}[ht]
    \centering
    \includegraphics[width=0.4\linewidth]{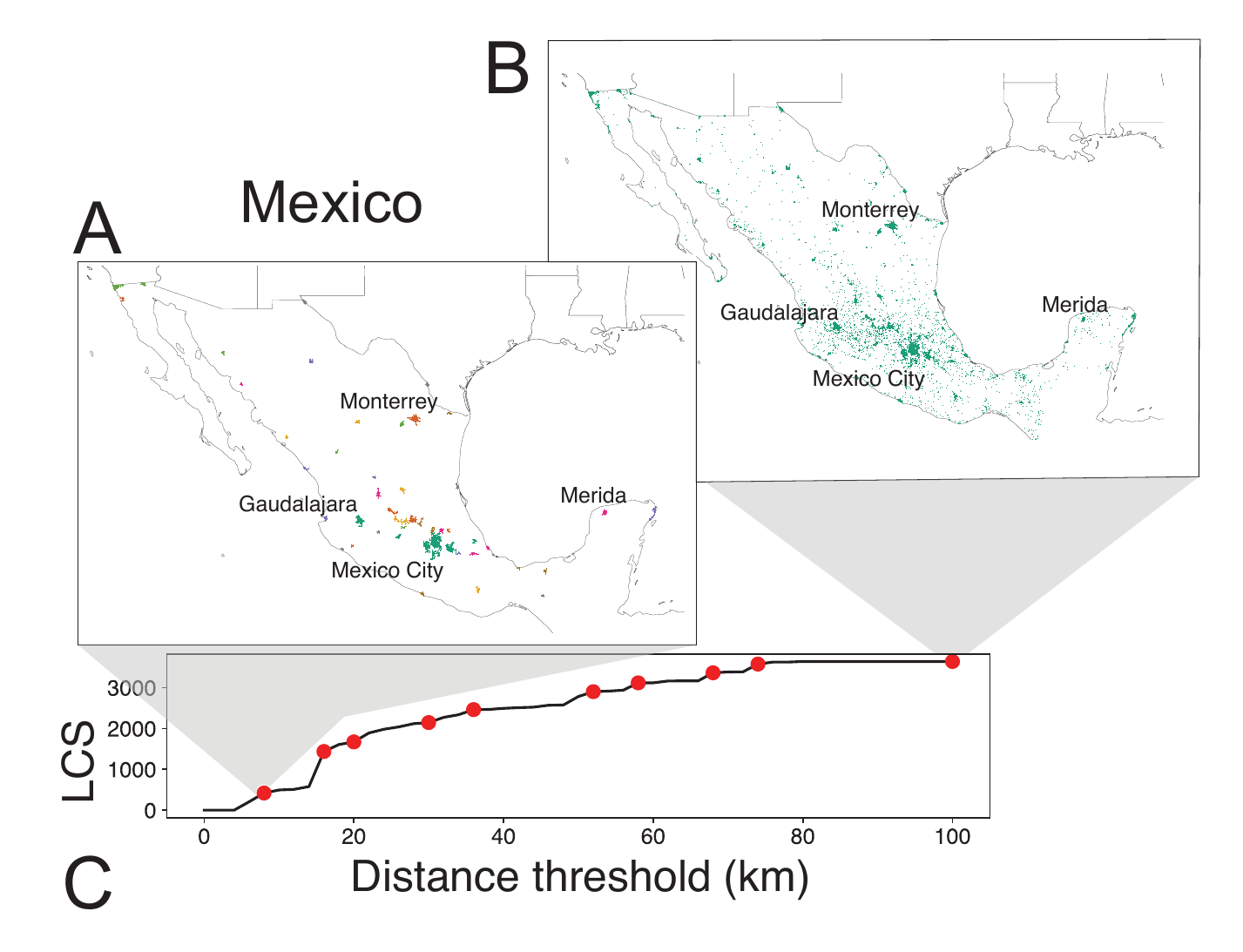}
    \caption{\textbf{Defining urban boundaries in Mexico by applying percolation to mobility networks.} \textbf{A-C)} Urban boundaries in Mexico at the first and last critical point in Mexico, respectively. \textbf{D)} Largest component size (LCS) at each distance threshold, with red points indicating each critical point in Mexico.}
    \label{fig:percolation_mx}
\end{figure}

\begin{figure}[!hb]
    \centering
    \includegraphics[width=0.4\linewidth]{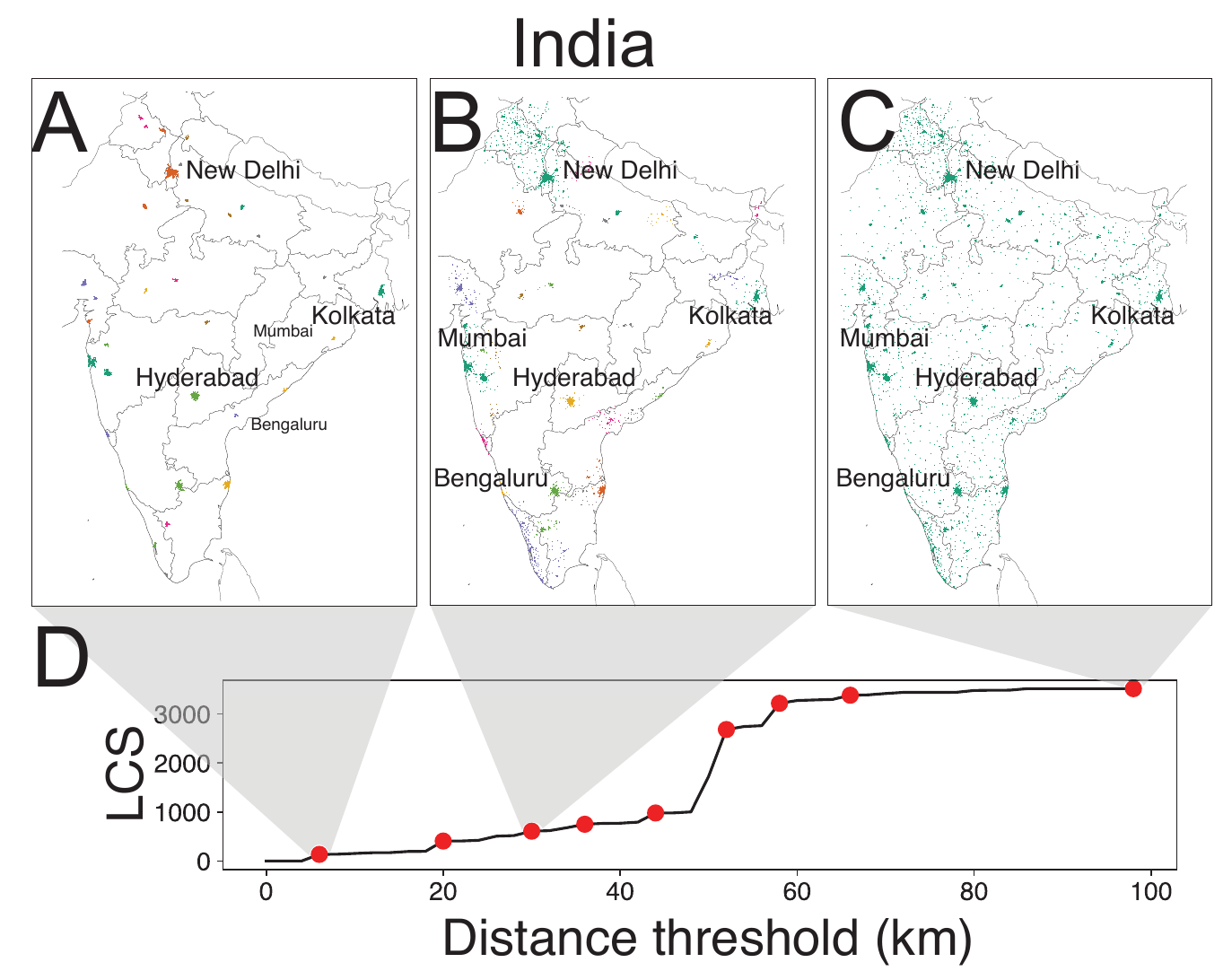}
    \caption{\textbf{Defining urban boundaries in India by applying percolation to mobility networks.} \textbf{A-C)} Urban boundaries in India at the first, third, and last critical point in India, respectively. \textbf{D)} Largest component size (LCS) at each distance threshold, with red points indicating each critical point in India.}
    \label{fig:percolation_in}
\end{figure}

Figures \ref{fig:percolation_mx}-\ref{fig:percolation_in} show the urban boundaries at the first critical point in Mexico and India, respectively. As discussed in the manuscript, these are derived by applying percolation to the mobility network. They are complementary to Figure 4 in the main manuscript, which shows the urban boundaries for Colombia. 

\newpage
\subsection{Validating Urban Rural Definitions}\label{ssec:validating_classification}

To validate our approach to defining urban and rural areas, we leverage the global urban and rural settlement (GURS) dataset \cite{liu2024global}, which distinguishes the two types of areas from 2000 to 2020, at a 100m spatial resolution, using geospatial data on the built-up surface, nighttime light, and place information from OpenStreetMap \cite{OpenStreetMap}. 
This dataset reflects an approach that uses static data to delineate urban and rural areas from one another. 
While traditional methods rely on density and built environment features to classify urban and rural areas, analysing mobility patterns offers a more nuanced and dynamic understanding since density alone may not capture the spatial relationship between places.
It is important to note the stark contrast in spatial resolutions, with the GURS dataset being measured at 100 metre resolution and our classification being based on geohash length 5 (approximately 5 kilometres). 

Some geohashes, such as those containing industrial zones or business parks, may have a higher density of built environment and low mobility activity leading to an urban classification using conventional methods, but a rural one, when leveraging mobility data. 
Alternatively, rural areas that are highly dependent on urban areas for resources such as employment or healthcare services may be classified as urban, despite having conventionally rural features. 
Table \ref{tab:class_validation} highlights the percentage of urban and rural validation data (columns) that exists within urban geohashes (third row) and rural geohashes (last row). We observe that Colombia and India both have more than 50\% of the urban validation data appearing in rural geohashes.

\begin{table}
\caption{\textbf{Percentage of the validation GURS data \cite{liu2024global} that aligns with our mobility-based classification.} Each column reflects urban and rural classifications from the GURS dataset. Meanwhile, rows reflect the urban and rural classification that we identify by applying network percolation on mobility data. The diagonal of each countries accuracy table reflect the true positive accuracy rate for urban and rural areas, respectively.}
\label{tab:class_validation}
\begin{tabular}{l|rr|rr|rr}
\toprule
& \multicolumn{6}{|c}{GURS Data} \\
\cmidrule(lr){2-7}
& \multicolumn{2}{|c}{Colombia} & \multicolumn{2}{|c}{Mexico} & \multicolumn{2}{|c}{India} \\
\cmidrule(lr){2-7}
 & Urban & Rural & Urban & Rural & Urban & Rural \\
\midrule
Mobility-based Urban & 49.06 & 24.48 & 70.36 & 45.36 & 44.07 & 20.06 \\
Mobility-based Rural & 50.94 & 75.52 & 29.64 & 54.64 & 55.93 & 79.94 \\
\bottomrule
\end{tabular}
\end{table}

\begin{figure}
    \centering
    \includegraphics[width=\linewidth]{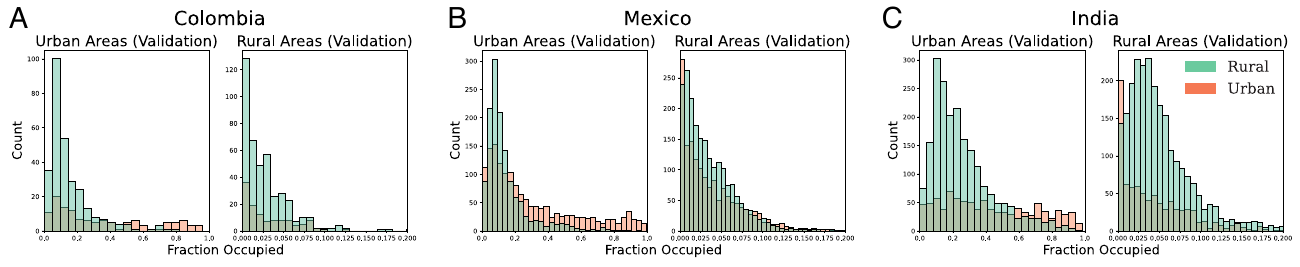}
    \caption{\textbf{Distribution of the fraction of a geohash that is present in the urban and rural validation data in Colombia (A), Mexico (B), and India (C).} The left and right panels of each country reflect the covered area for urban (green) and rural (orange) geohashes, considering the urban and rural validation data, respectively.}
    \label{si_fig:validatio_urbanRural_fraction}
\end{figure}

Given the distinct spatial scales between the two datasets, we can consider how much of a geohash intersects with urban and rural areas in the validation data. Figure \ref{si_fig:validatio_urbanRural_fraction} conveys how the difference in spatial resolution impacts accuracy by showing the fraction of a geohash that the urban and rural validation data makeup.

We calculate the total area of rural validation data in a geohash, finding that this value ranges from 0\% to 20\% of a geohash's area.
This is much lower than the total area of urban validation data in a geohash, which ranges close to full coverage for some geohashes. In the case of Mexico (left panels of Figure \ref{si_fig:validatio_urbanRural_fraction}B), we observe that when an urban area in the validation data intersects with a rural area in our classification, it tends to be a small level of intersection, typically not covering more than 50\% of the relevant geohash. This is not the case for Colombia and India (left panels of Figure \ref{si_fig:validatio_urbanRural_fraction}A and \ref{si_fig:validatio_urbanRural_fraction}C). This could indicate that Colombia and India tend to have built-up areas that exhibit low mobility to other areas. In this manner, conventional methods would identify this region as urban, while our mobility-based approach would consider it to be rural. 

\begin{figure}[!hb]
    \centering
    \includegraphics[width=0.9\linewidth]{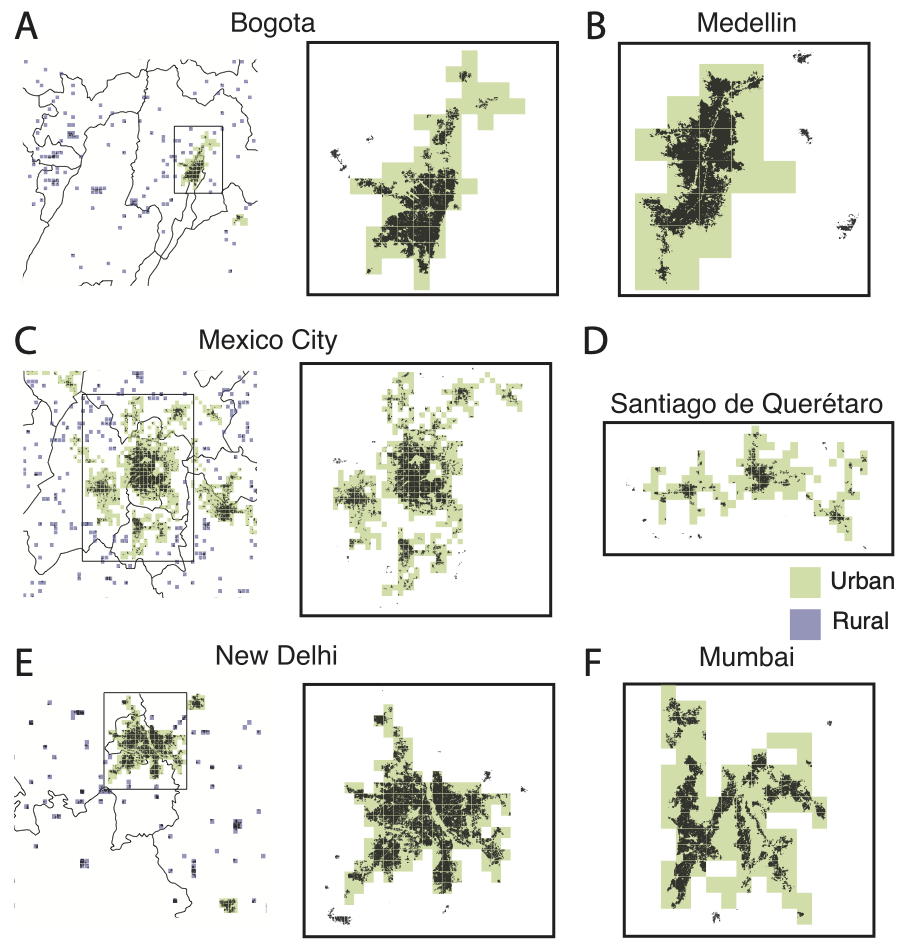}
    \caption{\textbf{Geovisualisation of the intersection between urban areas defined by conventional methods (i.e. built environment, satellite imagery) and our approach (applying network percolation on a mobility network).} The left panels of (A), (C), and (E), show the urban validation data (black) overlayed upon our classification, with green cells reflect urban geohashes and purple cells representing rural geohashes. Meanwhile the right panels of (A), (C), (E), and the (B), (D), (F) show how the different urban geohashes can exhibit various coverage areas and provides a visual example of how our approach, }
    \label{si_fig:validation_geo}
\end{figure}

This idea is better highlighted in the left panel of Figure \ref{si_fig:validation_geo}C, in which urban areas in the validation data (black cells) to the north-east of Mexico City are also classified as urban by our approach (green cells), but areas that are north-west and a similar distance away are categorised as rural (purple cells). 
Thus, the left panels of Figures \ref{si_fig:validation_geo}A, \ref{si_fig:validation_geo}C, and \ref{si_fig:validation_geo}E highlight where misalignment between our rural-urban definition and more conventional ones occur. Focusing on urban areas, the right panels of Figures Figures \ref{si_fig:validation_geo}A, \ref{si_fig:validation_geo}C, and \ref{si_fig:validation_geo}E, as well as Figures \ref{si_fig:validation_geo}B, \ref{si_fig:validation_geo}D, and \ref{si_fig:validation_geo}F shows how the geohashes in an urban area intersect with the validation urban data. 

Ultimately, cities and rural areas are not just defined by buildings but by how people interact with spaces. Since mobility data directly reflects human movement patterns, our approach provides a complementary and dynamic perspective to traditional, infrastructure-based classifications. Despite these inconsistencies, the classification presented in this work remains valuable as it captures urbanisation in a functional sense rather than a purely physical one.

\newpage
\subsection{Evaluating Biases in Mobility Data}\label{ssec:mobility_bias}

\begin{figure}[!hb]
    \centering
    \includegraphics[width=0.7\textwidth]{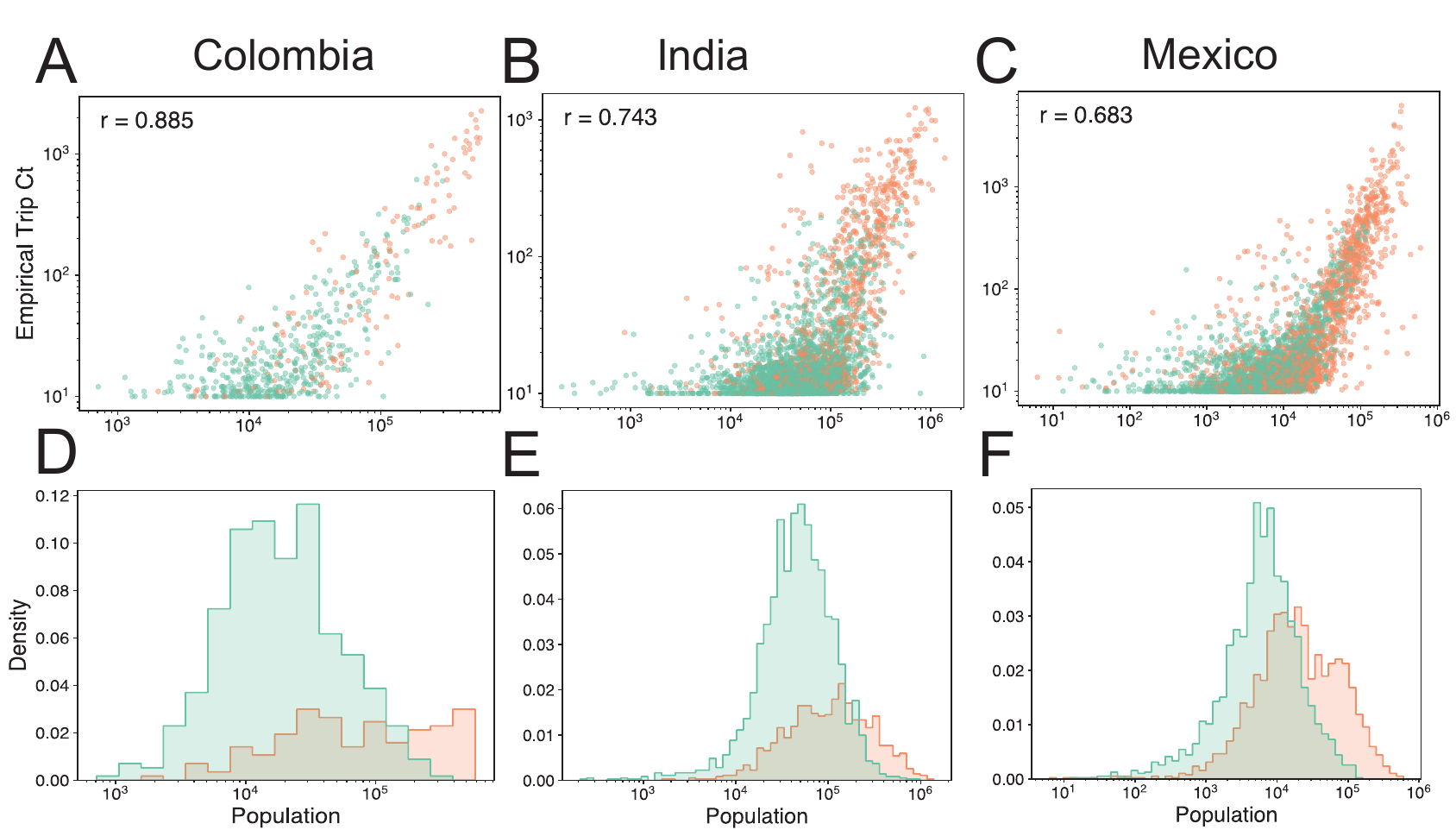}
    \caption{\textbf{Population characteristics for rural and urban areas in all three countries. A-C)} Relationship between empirical mobility trips and population in Colombia, India, and Mexico, respectively. \textbf{D-F)} Probability density function for the population in urban (orange) and rural (green) geohashes in Colombia, India, and Mexico, respectively.}
    \label{fig:pop_val}
\end{figure}

Mobility data, often sourced from mobile phones or transportation systems, represents movement trends but may not directly correlate with the number of individuals due to varying device ownership rates, demographic differences, or data collection biases. Comparing this data with population size helps contextualise findings, allowing us to identify whether observed activity levels reflect genuine population behaviour or are artefacts of uneven data coverage. We find that the empirical trip counts from each geohash, informed by the Spectus mobility data, are strongly correlated to the geohash population counts, provided by World Pop \cite{bondarenko2020census}  (Figure \ref{fig:pop_val}A-C). 

We detail population characteristics of urban and rural areas. Focusing on the mobility data provided by Spectus, Figures \ref{fig:pop_val}D-F shows the probability density function for geohashes' populations. Although disaggregated by urban (orange) and rural (green) areas, the normalisation is applied over the entire data, highlighting two main characteristics of our data and definitions of urban and rural: (1) there are less urban geohashes than rural and (2) urban geohashes tend to be more populated.

\newpage
\section{Supplementary note 2: Mobility Inefficiency Characteristics}

\subsection{Rural-Urban Mobility Inefficiency}\label{sec:mob_ineff_chars}
\begin{figure}[h]
    \centering
    \includegraphics[width=0.7\linewidth]{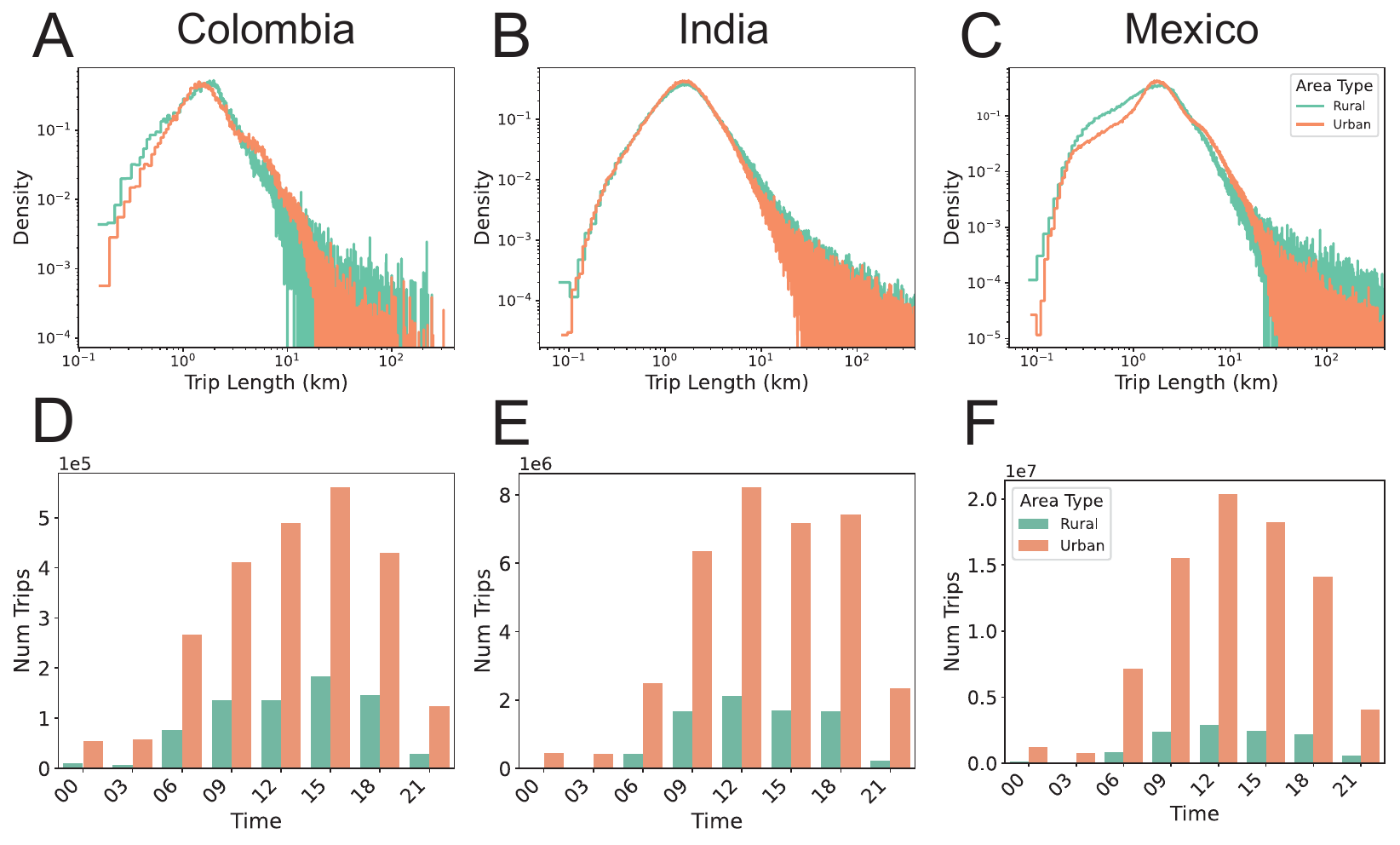}
    \caption{\textbf{Trip characteristics for rural and urban areas in all three countries. A-C)} Trip length distribution from urban (orange) and rural (green) geohashes in Colombia, India, and Mexico, respectively. \textbf{D-F)} Trip count distribution across different times in Colombia, India, and Mexico, respectively.}
    \label{fig:si_ur_trip_diff}
\end{figure}

 Here we show that the findings presented in the main manuscript are indeed a significant result. Table \ref{tab:ru_ineff_statDiff} highlights the statistical properties of the mobility inefficiency distribution across rural and urban areas in each of the three countries. We can observe that their differences do not exceed 470 metres, which is negligible given the range of inefficiency in the data. Figures \ref{fig:si_ur_trip_diff}A-C shows that trips shorter than 5km tend to occur more frequently than longer ones. Meanwhile, Figures \ref{fig:si_ur_trip_diff}D-F highlight that more trips tend to occur during 9AM-6PM. 
 
\begin{table}[]
    \centering
    \begin{tabular}{lrrrr}
    \toprule
     & \multicolumn{2}{c}{Mean} & \multicolumn{2}{c}{Median} \\
     \hline
    Country & Urban & Rural & Urban & Rural \\
    \midrule
    Colombia & -373.77 & 96.09 & -603.96 & -279.78 \\
    Mexico & -192.96 & 169.54 & -531.15 & -155.06 \\
    India & -100.69 & 310.73 & -473.99 & -141.20 \\
    \bottomrule
    \end{tabular}
    \caption{Statistical properties of urban and rural mobility inefficiency distribution}
    \label{tab:ru_ineff_statDiff}
\end{table}
\subsection{Distance and Time Dependent Inefficiency} \label{ssec:distanct_time_dependent_inefficiency}

\begin{figure}[h]
    \centering
    \includegraphics[width=0.7\linewidth]{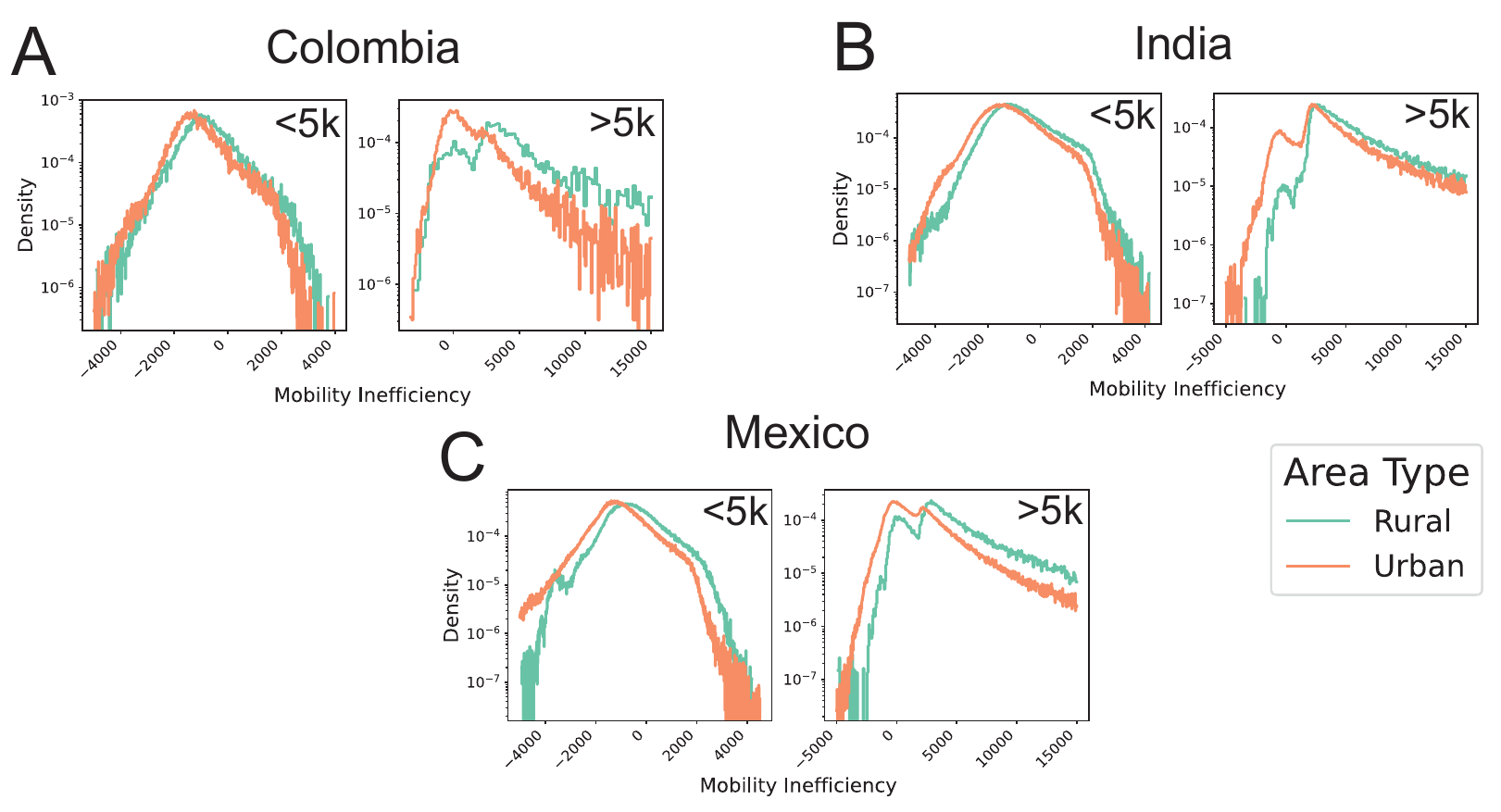}
    \caption{\textbf{Inefficiency characteristics for short and long trips in rural and urban areas in Colombia (A), India (B), and Mexico (C)} The left and right panels in each figure shows the inefficiency distribution for trips less than 5km and greater than 5km, respectively.}
    \label{fig:si_ur_mobIneff_dist}
\end{figure}

\begin{figure}[h]
    \centering
    \includegraphics[width=0.7\linewidth]{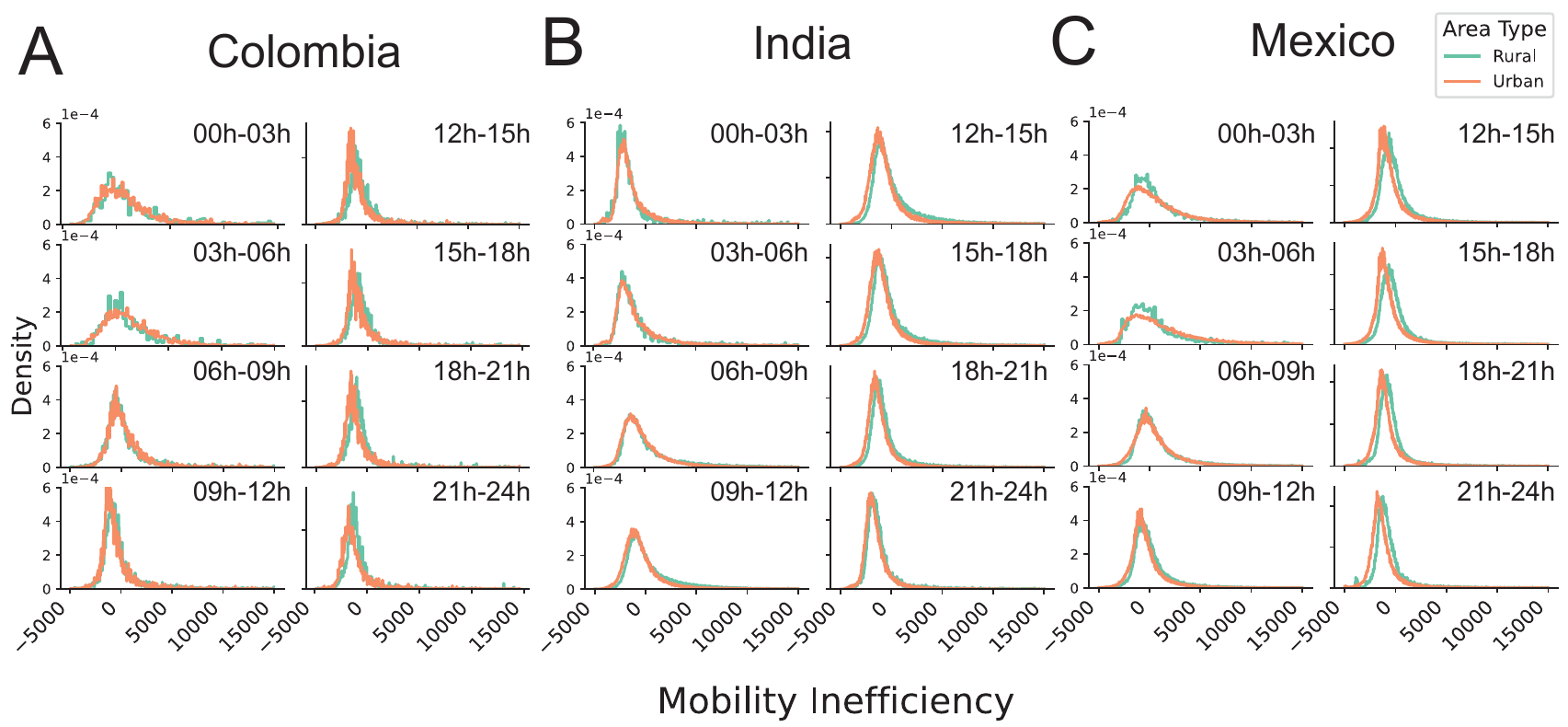}
    \caption{\textbf{Inefficiency distribution between rural and urban areas across 3-hour time intervals in Colombia (A), India (B), and Mexico (C)} Rural (green) and urban (orange) distributions throughout the day, highlight how mobility inefficiency changes throughout the day, yet the two area types show few discrepancies in their distributions}
    \label{fig:si_ur_mobIneff_temporal}
\end{figure}

With this in mind, we evaluate mobility inefficiency based on both trip length and time, highlighting that significant differences emerge only when these features are considered together in relation to inefficiency, rather than evaluating each feature individually. Specifically, Figures \ref{fig:si_ur_mobIneff_dist}A-C demonstrate how rural and urban areas typically experiences similar levels of inefficiency for shorter and longer trips in Colombia, India, and Mexico. Similarly, rural and urban areas tend to have similar inefficiency probability density distributions at different time intervals (Figure \ref{fig:si_ur_mobIneff_temporal}).

Finally, we provide the complementary plots for Figure \ref{fig:distance_inefficiency}B, showing how the trend of rural areas experiencing more inefficiency for longer trips that are later in the day are consistent for Mexico and India as well.  

\begin{figure}[h]
    \centering
    \includegraphics[width=0.7\linewidth]{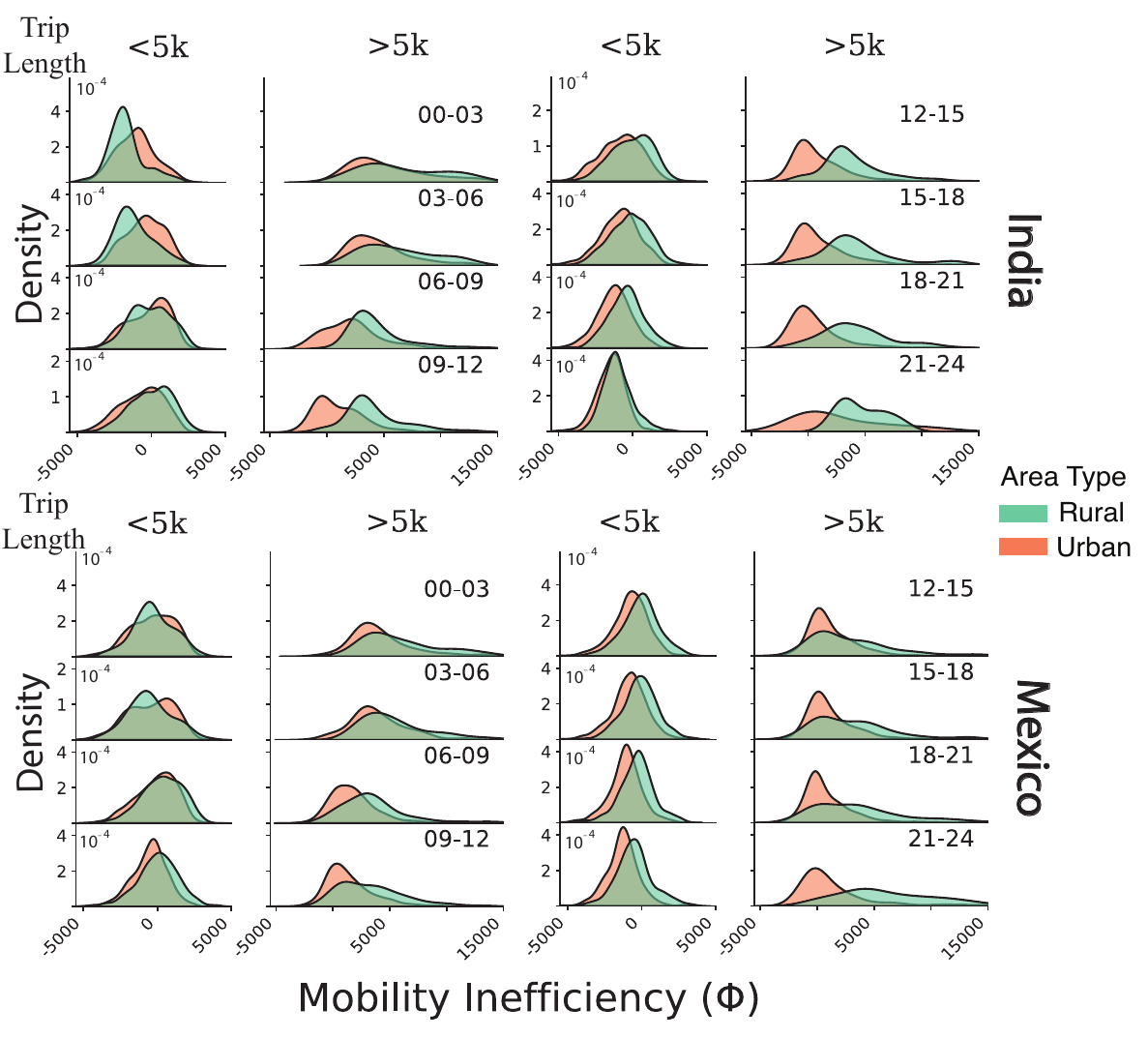}
    \caption{\textbf{Mobility inefficiency for short and long trips and different time intervals in India (top) and Mexico (bottom).} Rural (green) and urban (orange) distributions throughout the day and for short (<5km) and long (>5km) trips, highlight how rural areas tend to experience more inefficiency for longer trips, later in the day.}
    \label{fig:si_ur_mobIneff_temporalSpatial}
\end{figure}

\newpage \pagebreak \cleardoublepage
\section{Supplementary Note 3: Transit Infrastructure} 
\subsection{Transit Access in Rural and Urban Areas}\label{sec:transit_infrastructre}
\begin{figure}[h]
    \centering
    \includegraphics[width=0.9\linewidth]{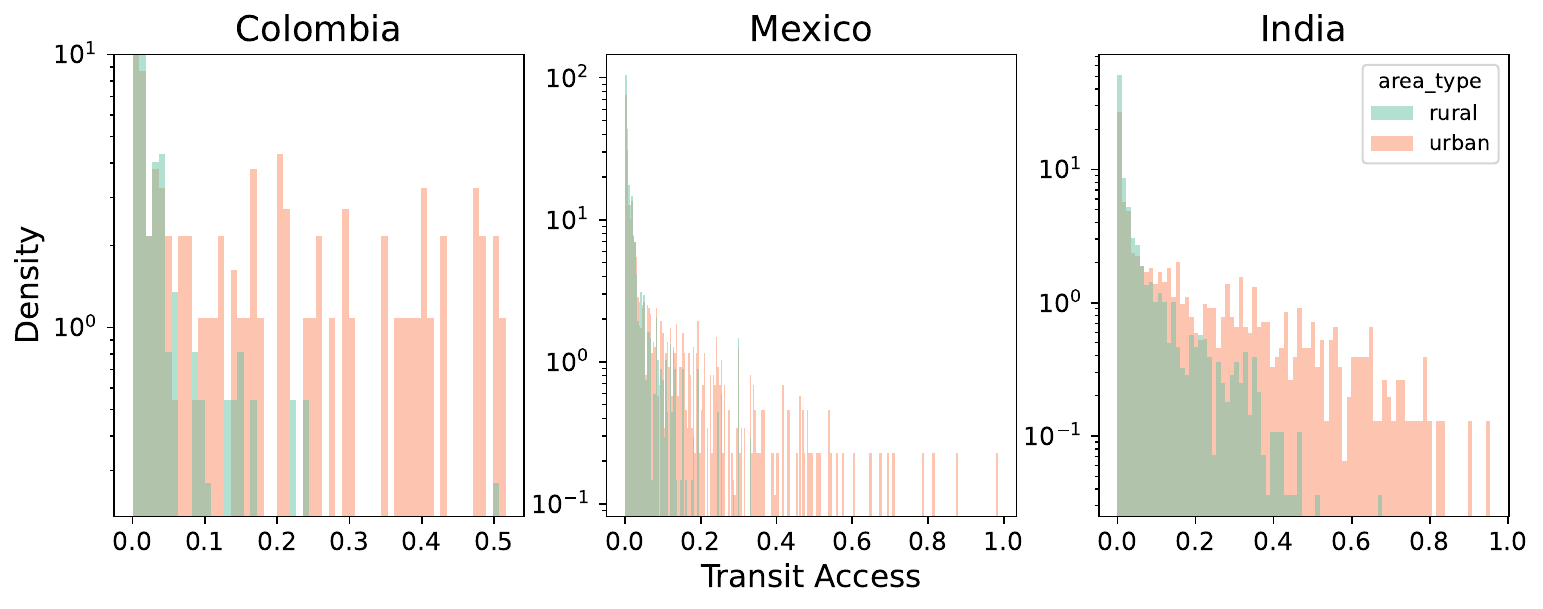}
    \caption{Caption}
    \label{fig:si_pt_score_distr}
\end{figure}

To provide further context to the transit infrastructure scores used in the main manuscript, we highlight the distribution of transit access across all urban geohashes in Colombia, Mexico, and India, respectively (\ref{fig:si_pt_score_distr}). 

Figure \ref{fig:si_pt_score_distr} emphasises how urban geohashes with higher transit access scores tend to be less prominent.

\newpage \pagebreak \cleardoublepage
\section{Supplementary Note 4: Vector Field Alternatives}

\subsection{A Local-Max Approach to Vector Fields}\label{sec:vector_field_alt}
Here, we present a simpler alternative to building transit access and mobility efficiency vector fields. The transit access vector field for a city is defined such that the direction of a vector points toward the local maximum infrastructure score (considering all 8 neighbours of a geohash), while the magnitude corresponds to the score of the focal geohash. Mathematically, the transit vector field can be defined as $\textbf{F}_T = (U_T, V_T)$, where $U_T$ and $V_T$ reflect the horizontal and vertical components of the vector field, respectively. Similarly, we can define the mobility efficiency vector field ($\textbf{F}_E = (U_E, V_E$) such that each geohash’s vector points toward the local minimum efficiency score among adjacent geohashes, and its magnitude represents the absolute value of the inefficiency ($\phi$) at the focal geohash. This means that the efficiency vector field is directed at geohashes with the most efficiency (i.e., lowest $\phi$ values).

Furthermore, for each city in Figure \ref{fig:infrastructure_inefficiency}D, we compare the sequence of efficiency transit alignment when using the mathematical formulation presented in the main manuscript to sequence when using the simpler vector fields, defined above. Table \ref{si_tab:validate_vector_field} shows strong positive correlations of alignment between the two methods, for each of the 15 cities.

\begin{table}[]
    \centering
    \begin{tabular}{ll}
        \toprule
        City & Correlation Coefficient (r) \\
        \midrule
        Mexico City & \textbf{0.731**} \\
        Santiago de Querétaro & \textbf{0.797***} \\
        Puebla & \textbf{0.695**} \\
        Monterrey & \textbf{0.876***} \\
        Guadalajara & \textbf{0.752***} \\
        New Delhi & \textbf{0.885***} \\
        Mumbai & \textbf{0.887***} \\
        Bengaluru & \textbf{0.922***} \\
        Hyderabad & \textbf{0.975***} \\
        Kolkata & \textbf{0.951***} \\
        Bogotá & \textbf{0.924***} \\
        Medellín & \textbf{0.823***} \\
        Cali & \textbf{0.903***} \\
        Barranquilla & \textbf{0.9***} \\
        Cartagena & \textbf{0.852***} \\
        \midrule
        \textbf{*}p<0.05, \textbf{**}p<0.01, \textbf{***}p<0.001 \\
    \end{tabular}
    \caption{Correlation of efficiency-transit alignment using two different methodologies to generate vector fields. }
    \label{si_tab:validate_vector_field}
\end{table}

\section{Supplementary Note 5: Rural Inefficiency in relation to Proximate Cities} 
\subsection{Understanding Rural Inefficiency and Proximity to Cities using OLS}\label{sec:ols_extra}
In this section we provide the results of applying OLS to the rural areas that surround 15 cities across Colombia (Table \ref{si_tab:co_rural_ineff}), India (Table \ref{si_tab:in_rural_ineff}) and Mexico (Table \ref{si_tab:mx_rural_ineff}). We show that mobility inefficiency in rural areas is highly contextual in nature, and that only rural areas that surround certain cities (i.e. New Delhi, Bogotá, Puebla, etc.) can be explained by features such as distance from the respective city, and their orientation around the city.

\begin{table}
\caption{OLS results for rural areas surrounding the 5 largest cities in Colombia (Cartagena, Barranquilla, Bogotá, Cali, Medellín), shown along the columns. Each row reflects the different features which are used to explain the target variable (mobility inefficiency). The coefficient of each feature, for each city is displayed in the corresponding cell, with standard errors in parentheses and asterisks reflect the significance of the respective coefficient.}
\label{si_tab:co_rural_ineff}
\centering
\begin{tabular}{llllll}
\hline
                     & Cartagena & Barranquilla & Bogota      & Cali        & Medellin    \\
\hline
const                & 2104.29   & -177.70      & \textbf{-3416.18***} & \textbf{-1932.39***} & \textbf{-2167.06**}  \\
                     & (2807.05) & (864.89)     & (1198.63)   & (620.89)    & (1000.31)   \\
Distance             & 3.51      & -0.53        & \textbf{21.63**}     & \textbf{4.97**}      & \textbf{10.87**}     \\
                     & (5.88)    & (3.38)       & (8.20)      & (2.16)      & (5.31)      \\
Population           & \textbf{0.01**}    & 0.00         & 0.01        & \textbf{0.01**}      & \textbf{0.01**}      \\
                     & (0.00)    & (0.01)       & (0.01)      & (0.00)      & (0.00)      \\
East-West Orientation     & 72.07     & 217.54       & 254.19      & \textbf{817.78**}    & 333.04      \\
                     & (988.61)  & (585.99)     & (427.20)    & (324.66)    & (376.41)    \\
North-South Orientation     & 4479.83   & -696.93      & \textbf{-1031.48**}  & 147.30      & 272.49      \\
                     & (2847.84) & (723.32)     & (506.08)    & (213.98)    & (268.37)    \\
Proximity to cities & 18.53     & 1.90         & \textbf{30.64*}      & \textbf{3.10*}       & 1.07        \\
                     & (11.42)   & (10.01)      & (15.49)     & (1.73)      & (3.85)      \\
R-squared            & 0.29      & 0.04         & 0.17        & 0.12        & 0.23        \\
\hline
Standard errors in parentheses. \\ 
\textbf{*} p<.1, \textbf{**} p<.05, \textbf{***}p<.01
\end{tabular}
\end{table}

\begin{table}
\caption{OLS results for rural areas surrounding the 5 largest cities in India (Mumbai, Kolkata, New Delhi, Bengaluru, Hyderabad), shown along the columns. Each row reflects the different features which are used to explain the target variable (mobility inefficiency). The coefficient of each feature, for each city is displayed in the corresponding cell, with standard errors in parentheses and asterisks reflect the significance of the respective coefficient.}
\label{si_tab:in_rural_ineff}
\begin{center}
\begin{tabular}{llllll}
\hline
                     & Mumbai    & Kolkata   & New Delhi  & Bengaluru & Hyderabad   \\
\hline
const                & -6400.68  & -965.33   & -1438.47   & -568.54   & \textbf{-2996.96**}  \\
                     & (6866.08) & (1916.94) & (1044.01)  & (1037.49) & (1323.63)   \\
Distance             & 64.51     & -1.45     & -1.29      & -0.68     & \textbf{10.48**}     \\
                     & (92.08)   & (6.42)    & (6.81)     & (5.50)    & (5.21)      \\
Population           & 0.02      & \textbf{0.01*}     & \textbf{0.03***}    & \textbf{0.01**}    & \textbf{0.01***}     \\
                     & (0.02)    & (0.00)    & (0.01)     & (0.00)    & (0.00)      \\
East-West Orientation    & 1437.64   & -144.37   & \textbf{-1662.90**} & -148.73   & \textbf{853.35**}    \\
                     & (1970.12) & (963.12)  & (730.53)   & (209.29)  & (383.58)    \\
North-West Orientation     & -401.51   & 618.21    & -542.27    & -6.08     & -128.77     \\
                     & (688.50)  & (480.65)  & (456.48)   & (402.46)  & (193.95)    \\
Proximity to cities & 24.15     & -1.99     & \textbf{-31.05**}   & 0.09      & \textbf{9.82**}      \\
                     & (53.78)   & (9.88)    & (13.93)    & (4.63)    & (4.83)      \\
R-squared            & 0.36      & 0.07      & 0.30       & 0.07      & 0.11        \\
\hline
Standard errors in parentheses. \\ 
\textbf{*} p<.1, \textbf{**} p<.05, \textbf{***}p<.01
\end{tabular}
\end{center}
\end{table}

\begin{table}
\caption{OLS results for rural areas surrounding the 5 largest cities in Mexico (Mexico City, Guadalajara, Puebla, Monterrey, Querétaro), shown along the columns. Each row reflects the different features which are used to explain the target variable (mobility inefficiency). The coefficient of each feature, for each city is displayed in the corresponding cell, with standard errors in parentheses and asterisks reflect the significance of the respective coefficient.}
\label{si_tab:mx_rural_ineff}
\begin{center}
\begin{tabular}{llllll}
\hline
                     & Mexico City & Guadalajara & Puebla     & Monterrey & Santiago de Querétaro  \\
\hline
const                & 18307.81    & -2205.07    & \textbf{-1952.28**} & 2074.72   & 180.82                 \\
                     & (34673.32)  & (1360.42)   & (762.34)   & (1846.91) & (1345.72)              \\
Distance             & 18.59       & 16.19       & \textbf{16.53**}    & -10.30    & -11.46                 \\
                     & (71.13)     & (17.23)     & (7.73)     & (10.37)   & (15.47)                \\
Population           & 0.03        & \textbf{0.05**}      & \textbf{0.05**}     & 0.05      & 0.04                   \\
                     & (0.03)      & (0.02)      & (0.02)     & (0.03)    & (0.03)                 \\
East-West Orientation    & 11845.08    & -16.26      & \textbf{763.00**}   & -98.41    & \textbf{1972.94*}               \\
                     & (12453.74)  & (565.47)    & (305.39)   & (818.51)  & (1083.15)              \\
North-South Orientation     & 19762.95    & -442.91     & -269.80    & 472.10    & 343.54                 \\
                     & (44746.13)  & (546.05)    & (475.34)   & (625.05)  & (693.42)               \\
Proximity to cities & 52.63       & 31.10       & -6.32      & -31.22    & -36.44                 \\
                     & (252.85)    & (23.97)     & (10.57)    & (31.24)   & (24.44)                \\
R-squared            & 0.37        & 0.28        & 0.31       & 0.24      & 0.26                   \\
\hline
Standard errors in parentheses. \\ 
\textbf{*} p<.1, \textbf{**} p<.05, \textbf{***}p<.01
\end{tabular}
\end{center}
\end{table}
